\documentclass[%
 reprint,
 amsmath,amssymb,
 aps,
]{revtex4-2}

\usepackage{graphicx}
\usepackage{dcolumn}
\usepackage{bm}
\usepackage{hyperref}
\hypersetup{
    colorlinks=true,
    linkcolor=blue,
    urlcolor=blue,
    citecolor=blue
}
\usepackage[mathlines]{lineno}

\usepackage{multirow} 
\usepackage{booktabs}

\begin{document}

\preprint{APS/123-QED}

\title{Measurements of the cross sections of $e^{+}e^{-}\to \Sigma_{c}\bar{\Sigma}_{c}$ and  $\Lambda_c^{+}\bar{\Sigma}_{c}^{-}$ near kinematic thresholds}

\author{
\begin{small}
\begin{center}
\begin{small}
\begin{center}
M.~Ablikim$^{1}$, M.~N.~Achasov$^{4,c}$, P.~Adlarson$^{77}$, X.~C.~Ai$^{82}$, R.~Aliberti$^{36}$, A.~Amoroso$^{76A,76C}$, Q.~An$^{73,59,a}$, Y.~Bai$^{58}$, O.~Bakina$^{37}$, Y.~Ban$^{47,h}$
% ...
\\
\vspace{0.2cm}
(BESIII Collaboration)\\
\vspace{0.2cm} {\it
$^{1}$ Institute of High Energy Physics, Beijing 100049, People's Republic of China\\
% ...
$^{4}$ Budker Institute of Nuclear Physics SB RAS (BINP), Novosibirsk 630090, Russia\\
% ...
$^{36}$ Johannes Gutenberg University of Mainz, Johann-Joachim-Becher-Weg 45, D-55099 Mainz, Germany\\
$^{37}$ Joint Institute for Nuclear Research, 141980 Dubna, Moscow region, Russia\\
% ...
$^{47}$ Peking University, Beijing 100871, People's Republic of China\\
% ...
$^{58}$ Southeast University, Nanjing 211100, People's Republic of China\\
$^{59}$ State Key Laboratory of Particle Detection and Electronics, Beijing 100049, Hefei 230026, People's Republic of China\\
% ...
$^{73}$ University of Science and Technology of China, Hefei 230026, People's Republic of China\\
% ...
$^{76}$ University of Turin and INFN, (A)University of Turin, I-10125, Turin, Italy; (B)University of Eastern Piedmont, I-15121, Alessandria, Italy; (C)INFN, I-10125, Turin, Italy\\
$^{77}$ Uppsala University, Box 516, SE-75120 Uppsala, Sweden\\
% ...
$^{82}$ Zhengzhou University, Zhengzhou 450001, People's Republic of China\\
\vspace{0.2cm}
$^{a}$ Deceased\\
% ...
$^{c}$ Also at the Novosibirsk State University, Novosibirsk, 630090, Russia\\
% ...
$^{h}$ Also at State Key Laboratory of Nuclear Physics and Technology, Peking University, Beijing 100871, People's Republic of China\\
% ...
}\end{center}

\vspace{0.4cm}
\end{small}
\end{center}
\end{small}
}

\date{\today}

\begin{abstract}
Based on $e^{+}e^{-}$ annihilation data collected with the BESIII detector at the BEPCII collider, searches for the $e^{+}e^{-}\to\Sigma_{c}\bar{\Sigma}_{c}$ ($\Sigma_{c}^{+}\bar{\Sigma}_{c}^{-}$, $\Sigma_{c}^{0}\bar{\Sigma}_{c}^{0}$ and $\Sigma_{c}^{++}\bar{\Sigma}_{c}^{--}$) processes,  as well as for the isospin-violating process $e^{+}e^{-}\to\Lambda_c^{+}\bar{\Sigma}_{c}^{-}$, are conducted for the first time at $\sqrt{s}=4.750$, $4.781$, $4.843$, $4.918$, and $4.951$ GeV. As no signals are observed for all the above channels, the upper limits on their Born cross sections at the 90$\%$ confidence level are reported.
\end{abstract}

\maketitle


\section{\label{sec:intro}Introduction}

In the quark model~\cite{cit:quark_model}, baryons are not point-like particles but are described as being composed of three quarks. Although this model offers a valuable framework for understanding baryon structures, unresolved issues, such as the significant discrepancy between the proton mass and the sum of its three constituent quark masses~\cite{cit:Proton_size}, call for a critical re-examination of the quark model and highlight the need to better understand the interactions between the quarks inside the baryon.

The baryon properties can be indirectly examined through electron-positron collisions, where the two particles annihilate into a virtual photon, which converts into a baryon pair. The production cross section is parametrized in terms of electromagnetic form factors or effective form factors~\cite{cit:FF_concept}. Hence, the measurement of the production cross sections of baryon pairs near the mass threshold provides valuable insights into the internal structure of baryons~\cite{cit:FF_definition}.

The production cross sections and (effective) form factors have been studied in different experiments~\cite{cit:xs_exp_1, cit:xs_exp_2, cit:xs_exp_3_pp_xs, cit:xs_exp_4, cit:xs_exp_5,cit:nn_xs,cit:SND_nn_2022wdb,cit:SND_nn_2023fos,cit:SND_nn_2024pbk}, revealing several unexpected behaviours. 
For instance, in the hyperon sector, the cross sections of $e^{+}e^{-}\to\Sigma^{-}\bar{\Sigma}^{+}$, $e^{+}e^{-}\to\Sigma^{+}\bar{\Sigma^{-}}$, and $e^{+}e^{-}\to\Sigma^{0}\bar{\Sigma^{0}}$ are well described by perturbative Quantum Chromodynamics (pQCD) and follow a power law as a function of the energy~\cite{cit:sigmapsigmam_xs,cit:sigma0sigma0_xs}, while anomalous behaviors are observed for the $\Lambda\bar{\Lambda}$~\cite{cit:lmdlmd_xs} and $\Lambda\bar{\Sigma}^{0}$~\cite{cit:lmdsigma0_xs} cross sections, which differ from the pQCD predictions at threshold. Specifically, a threshold enhancement larger than what predicted by pQCD is observed for $\Lambda\bar{\Lambda}$ and $\Lambda\bar{\Sigma}^{0}$. 

With respect to the cross sections of charmed baryons, the Belle collaboration initially reported a resonant structure consistent with the $Y(4660)$ state near the $\Lambda_{c}^{+}\bar{\Lambda}_{c}^{-}$ mass threshold~\cite{cit:Belle_Y_LcLc_xs}. In contrast, the BESIII collaboration conducted a precise measurement of the cross sections of $e^{+}e^{-}\to\Lambda_{c}^{+}\bar{\Lambda}_{c}^{-}$ from threshold to 4.951 GeV, observed a threshold enhancement and extracted the corresponding form factors~\cite{cit:Bes_LcLc_xs_old,cit:Bes_LcLc_xs_new}, but with no indication of the resonant structure~\cite{cit:Bes_LcLc_xs_new}.


 
The study of the production cross sections of isospin-one charmed baryon ${\Sigma}_{c}$ pairs is of particular interest.
Firstly, the ${\Sigma}_{c}$ baryon, including the three isospin partners ${\Sigma}_{c}^{0}$ ($ddc$), ${\Sigma}_{c}^+$ ($udc$) and ${\Sigma}_{c}^{++}$ ($uuc$), has a considerable mass difference with respect to the $\Lambda_{c}^{+}$ ($udc$)~\cite{cit:ScSc_Mass_measurement_1,cit:ScSc_Mass_measurement_2}, even though they present similar quark components. This mass difference can be interpreted as the energy difference between the heavier ``bad" diquark $(ud)_{{\rm spin}=1}$, which is symmetric in light quark spin, and the lighter ``good" diquark $(ud)_{{\rm spin}=0}$, which is antisymmetric~\cite{cit:diquark_model}. Considering the ${\Lambda_{c}^{+}\bar{\Lambda}_{c}^{-}}$ and ${\Sigma_{c}^{+}\bar{\Sigma}_{c}^{-}}$ production, if the hadronization from the $c\bar{c}$ pair proceeds via attaching diquarks ($ud$), the cross section of the ${e^{+}e^{-}\to\Lambda_{c}^{+}\bar{\Lambda}_{c}^{-}}$ process is greater than the ${e^{+}e^{-}\to\Sigma_{c}^{+}\bar{\Sigma}_{c}^{-}}$ one, since the ``bad" diquark is heavier than the ``good" one, and the cross section is inversely proportional to the mass. Therefore, measuring ${\Sigma}^{+}_{c}$ pair cross section is crucial for understanding the characteristics of ``bad" and ``good" diquarks, thereby to get insight of the inner structure of charmed baryons.

Secondly, it is of interest to explore whether the ${\Sigma}_{c}$ pair can form a novel bound state, namely a dibaryon. Theories have various explanations for ${\Sigma}_{c}-\bar{\Sigma}_{c}$ interaction and have made predictions about potential new molecular states from both the hadron level and quark level. On the hadron level, several coupled channel effects, such as the one-pion-exchange model~\cite{cit:one_pion_exchange}, the one-boson-exchange model~\cite{cit:one_boson_exchange_1,cit:one_boson_exchange_2}, single channel Bethe-Salpeter equation~\cite{cit:one_coupling_channel}, and the chiral perturbation theory~\cite{cit:p_chiral_1,cit:p_chiral_2} have been developed. On the quark level, theoretical approaches like the chiral quark model~\cite{cit:chiral_quark}, quark delocalization color screening model~\cite{cit:quark_delocalazation}, and lattice QCD~\cite{cit:lattice_QCD} are employed to investigate bound state production mechanisms. Despite the wealth of predictions, the mechanisms influencing the ${\Sigma}_{c}-\bar{\Sigma}_{c}$ interaction remain inadequately explored in the absence of experimental data, and it is essential to conduct relevant measurements to validate these predictions.

The hadron-hadron interactions and the potential dibaryon for the 
$\Lambda_{c}^{+}\bar{\Sigma}^{-}_{c}$ system are also under discussion~\cite{cit:one_boson_exchange_2,cit:quark_delocalazation,cit:LcSc_theory_1}, and can be tested in the isospin violation process
${e^{+}e^{-}\to \Lambda_{c}^{+}\bar{\Sigma}^{-}_{c}}$, whose experimental studies are lacking. Moreover, isospin violation is expected to manifest via $e^+e^-\to \gamma^*\to q\bar{q}$ ($q$ denotes $u$, $d$ and $s$) in electromagnetic interaction. That raises concern regarding the relative amplitudes between isospin conversing and violating processes in charmed baryon production. In the isospin violating process $\gamma^*\to q\bar{q}$, under the diquark model, the diquark is initially produced and subsequently attaches to $c$ or $\bar{c}$ during the hadronization process, resulting in a suppressed cross section compared to the ${e^{+}e^{-}\to\Lambda_{c}^{+}\bar{\Lambda}^{-}_{c}}$ process. Thus, the measurement of the cross section ratio between ${e^{+}e^{-}\to\Lambda_{c}^{+}\bar{\Lambda}^{-}_{c}}$ and ${e^{+}e^{-}\to\Lambda_{c}^{+}\bar{\Sigma}^{-}_{c}}$ can shed light on the dynamics of the charmed baryon production and help to probe their inner structure from an alternative perspective.

In this paper, the aim is to measure the cross sections of $e^{+}e^{-}\to\Sigma_{c}\bar{\Sigma}_{c}$ ($\Sigma_{c}\bar{\Sigma}_{c}$ denotes $\Sigma_{c}^{+}\bar{\Sigma}_{c}^{-}$, $\Sigma_{c}^{0}\bar{\Sigma}_{c}^{0}$, or $\Sigma_{c}^{++}\bar{\Sigma}_{c}^{--}$) and $e^{+}e^{-}\to\Lambda_{\it c}^{+}\bar{\Sigma}_{c}^{-}$ in the energy regions near their mass thresholds. The analyzed data samples were collected with the BESIII detector at BEPCII collider. The data samples taken at $\sqrt{s} =4.918$ and $4.951~\mathrm{GeV}$, are utilized to measure the $e^{+}e^{-}\to\Sigma_{c}\bar{\Sigma}_{c}$ cross section. Subsequently, these data, along with those at $\sqrt{s} =4.750$, $4.781$, and $4.843~\mathrm{GeV}$, are utilized to conduct the latter measurement. The center-of-mass (CM) energies and integrated luminosities of these samples are measured from ${e^{+}e^{-}\to\Lambda_{c}^{+}\bar{\Lambda}_{c}^{-}}$ and Bhabha events, respectively~\cite{cit:4750_4946_data_lumin}, with the specific values shown in Table~\ref{tab:data_lum}. Four $\Lambda^{+}_{c}$ Cabibbo-favored decay modes, namely, $\Lambda_{c}^{+}\to p K_{S}^{0}, \Lambda_{c}^{+}\to pK^{-}\pi^{+}, \Lambda_{c}^{+}\to \Lambda\pi^{+}, {\rm and}~\Lambda_{c}^{+}\to \Sigma^{0}\pi^{+}$, are employed to reconstruct the $\Lambda^{+}_{c}$, while the $\Sigma_{c}$ is reconstructed by combining one residual pion with one $\Lambda^{+}_{c}$. Unless otherwise specified, charge-conjugate modes are always included.

\begin{table}[htbp]
\centering
\caption{ \label{tab:data_lum} Summary of the CM energy and integrated luminosities of the data samples. All values are from Ref.~\cite{cit:4750_4946_data_lumin}, where the first uncertainty represents statistical uncertainty and the second one represents systematic uncertainty.}
\begin{ruledtabular}
\begin{tabular}{cc}
$\sqrt{s}$~(MeV) & Luminosity~($\rm pb^{-1}$) \\ \hline
4750.05 $\pm$ 0.12 $\pm$ 0.29 & 366.55 $\pm$ 0.10 $\pm$ 1.95 \\
4780.54 $\pm$ 0.12 $\pm$ 0.30 & 511.47 $\pm$ 0.12 $\pm$ 2.72 \\
4843.07 $\pm$ 0.30 $\pm$ 0.32 & 525.16 $\pm$ 0.12 $\pm$ 2.79 \\
4918.02 $\pm$ 0.34 $\pm$ 0.34 & 207.82 $\pm$ 0.08 $\pm$ 1.10 \\
4950.93 $\pm$ 0.36 $\pm$ 0.38 & 159.28 $\pm$ 0.07 $\pm$ 0.85 
\end{tabular}
\end{ruledtabular}
\end{table}

The measurement of the $e^{+}e^{-}\to\Sigma_{c}\bar{\Sigma}_{c}$ cross section encounters several challenges. Firstly, it is difficult to disentangle the substantial contributions from $e^{+}e^{-}\to\Lambda^{+}_{c}\bar{\Lambda}_{c}^{*-}(2595)$ and $e^{+}e^{-}\to\Lambda^{+}_{c}\bar{\Lambda}_{c}^{*-}(2625)$ to the signal candidates $\Sigma_{c}\Sigma_{c}$, since they have overlapping kinematics and the same final states. Additionally, not only the two background channels $e^{+}e^{-}\to\Lambda^{+}_{c}\bar{\Lambda}_{c}^{*-}$ have significant contributions~\cite{cit:Lc2625_polar}, but also the decay rates of $\Lambda_{c}^{*+}\to \Lambda_{c}^{+} \pi\pi$ have large uncertainties~\cite{cit:PDG_value}. To overcome these challenges, three types of data samples are selected, including one sample of singly tagged $\Lambda^{+}_{c}$~\cite{cit:ST_defi},  three samples of  reconstructing one $\Lambda^{+}_{c}$ along one $\pi$ in three different charges, and two samples of combining a $\pi^+\pi^-$ or $\pi^0\pi^0$ with a $\Lambda^{+}_{c}$. To simplify, these strategies are abbreviated as ``Tag $\Lambda^{+}_{c}$", ``Tag $\Lambda^{+}_{c}\pi$", and ``Tag $\Lambda^{+}_{c}\pi\pi$", respectively, throughout this paper. A simultaneous fit is carried out to the three types of samples to determine the cross sections of the signal channels.
Furthermore, the ``Tag $\Lambda^{+}_{c}$" sample is utilized to measure the cross section of the $e^{+}e^{-}\to\Lambda_{c}^{+}\bar{\Sigma}_{c}^{-}$ process.

\section{\label{sec:dataset}Detector and data sample}

The BESIII detector~\cite{cit:BESIII_detector} records symmetric $e^+e^-$ collisions provided by the BEPCII storage ring~\cite{cit:BEPC}, which operates in the CM energy range from 1.84 to 4.95~GeV. The cylindrical core of the BESIII detector covers 93\% of the full solid angle and consists of a helium-based multilayer drift chamber~(MDC), a plastic scintillator time-of-flight system~(TOF), and a CsI(Tl) electromagnetic calorimeter~(EMC), which are all enclosed in a superconducting solenoidal magnet providing a 1.0~T magnetic field. The solenoid is supported by an octagonal flux-return yoke with resistive plate counter muon identification modules interleaved with steel. The charged-particle momentum resolution at $1~{\rm GeV}/c$ is $0.5\%$, and the ${\rm d}E/{\rm d}x$ resolution is $6\%$ for electrons from Bhabha scattering. The EMC measures photon energies with a resolution of $2.5\%$ ($5\%$) at $1$~GeV in the barrel (end cap) region. The time resolution in the TOF barrel region is 68~ps, while that in the end cap region is 110~ps. In 2015, the endcap TOF was replaced with a Multigap Resistive Plate Chamber, and the time resolution improved to 60~ps. All datasets mentioned in this paper benefit from this upgrade.

Monte Carlo (MC) simulated data samples produced with a {\sc geant4}-based~\cite{cit:GEANT4} software package, which includes the geometric description of the BESIII detector and the detector response, are used to determine detection efficiencies and to estimate backgrounds. The simulation models the beam energy spread and initial state radiation (ISR) in the $e^+e^-$
annihilations with the generator {\sc kkmc}~\cite{cit:KKMC,cit:KKMC_2}. All particle decays are modelled with {\sc evtgen}~\cite{cit:EVTGEN,cit:EVTGEN_2} using branching fractions either taken from the Particle Data Group~\cite{cit:PDG_value}, when available,
or otherwise estimated with {\sc lundcharm}~\cite{cit:lundcharm_1,cit:lundcharm_2}. Final state radiation from charged final state particles is incorporated using the {\sc photos} package~\cite{cit:PHOTOS}.

The $\Lambda_{c}^{+}$ events in signal MC samples are generated according to the partial wave analysis result of $\Lambda_{c}^{+}\to p K^{-}\pi^{+}$~\cite{cit:Lc2PKPi_PWA} and considering the polarization effects seen in data for the $\Lambda_{c}^{+}\to p K_{S}^{0}$, $\Lambda_{c}^{+}\to \Lambda\pi^{+}$, and $\Lambda_{c}^{+}\to \Sigma^{0}\pi^{+}$ channels. The $\Sigma_{c}\to\Lambda^{+}_{c}\pi$ is assumed as the only decay mode of $\Sigma_{c}$. 

The primary background sources include $e^{+}e^{-}\to\Lambda^{+}_{c}\bar{\Lambda}_{c}^{*-}(2595)$, $e^{+}e^{-}\to\Lambda^{+}_{c}\bar{\Lambda}_{c}^{*-}(2625)$, $e^{+}e^{-}\to\Lambda^{+}_{c}\bar{\Sigma}_{c}\pi$ ($\Lambda^{+}_{c}\bar{\Sigma}^{0}_{c}\pi^{-}$, $\Lambda^{+}_{c}\bar{\Sigma}^{--}_{c}\pi^{+}$, $\Lambda^{+}_{c}\bar{\Sigma}^{-}_{c}\pi^{0}$), $e^{+}e^{-}\to\Lambda^{+}_{c}\bar{\Lambda}_{c}^{-}$, and hadronic backgrounds. To estimate their contributions, the corresponding exclusive MC samples are also generated. For $\Lambda_{c}^{*+}(2595)$ and $\Lambda_{c}^{*+}(2625)$, only the strong decays, specifically $\Lambda_{c}^{*+}\to \Sigma_{c}^{0}\pi^{+}$, $\Sigma_{c}^{+}\pi^{0}$, $\Sigma_{c}^{++}\pi^{-}$, $\Lambda_{c}^{+}\pi^{+}\pi^{-}$, and $\Lambda_{c}^{+}\pi^{0}\pi^{0}$, are considered in the simulation with uniform kinematic distributions. Each of the five $\Lambda_{c}^{*+}$ decay channels is simulated in equal proportions and the relative ratio among the different decay channels is adjusted for further studies. For other hadronic backgrounds, the hadronic inclusive MC sample is simulated, including $D_{(s)}^{(*)}$ pair production, ISR return to the charmonium and charmonium-like $\psi$ states at lower masses, and continuum processes $e^{+}e^{-}\to q\bar{q}$.

\section{\label{sec:pre_selection}$\Lambda^{+}_{c}$ reconstruction}


The $\Lambda^{+}_{c}$  reconstruction is performed through final states characterized by $p$, $K_{S}^{0}(\to\pi^{+}\pi^{-})$, $K$, $\pi$, $\Lambda(\to p \pi^{-})$, and $\Sigma^{0}(\to\Lambda\gamma)$ particles.

Charged tracks detected in the MDC are required to be within a polar angle ($\theta$) range of $|\rm{cos\theta}|<0.93$, where $\theta$ is defined with respect to the $z$-axis,
which is the symmetry axis of the MDC. For charged tracks not originating from $K_S^0$, $\Lambda$ or $\Sigma^{0}$ decays, the distance of closest approach to the interaction point (IP) 
must be less than 10\,cm along the $z$-axis, $|V_{z}|$, and less than 1\,cm in the transverse plane, $|V_{xy}|$.


Particle identification~(PID) for charged tracks combines measurements of the energy deposited in the MDC~(d$E$/d$x$) and the flight time in the TOF to form likelihoods $\mathcal{L}(h)~(h=p,K,\pi)$ for each hadron $h$ hypothesis.
Tracks are identified as protons when the proton hypothesis has the greatest likelihood ($\mathcal{L}(p)>\mathcal{L}(K)$ and $\mathcal{L}(p)>\mathcal{L}(\pi)$), while charged kaons and pions are identified by comparing the likelihoods for the kaon and pion hypotheses, $\mathcal{L}(K)>\mathcal{L}(\pi)$ and $\mathcal{L}(\pi)>\mathcal{L}(K)$, respectively. These criteria are not applied for tracks used in reconstructing $K_{S}^{0}$ and to  the $\pi^{-}$ candidates for the $\Lambda$ decay.


Photon candidates are identified using isolated showers in the EMC. The deposited energy of each shower must be greater than $25$ MeV in the barrel region ($|\rm{cos}\theta| < 0.80$) and more than $50$ MeV in the end-cap region ($0.86 < |\rm{cos}\theta| < 0.92$). To suppress electronic noise and showers unrelated to the event, the difference between the EMC time and the event start time is required to be within [0, 700]\,ns.

The $K_{S}^{0}$ and $\Lambda$ candidates  are reconstructed from two oppositely charged tracks satisfying $|V_{z}|<$ 20~cm, the $\pi^{+}\pi^{-}$ and $p \pi^{-}$ final states, respectively. The two tracks are constrained to originate from a common decay vertex, with a requirement on the $\chi^{2}$ of the vertex fit being less than 100. Additionally, the decay length of the candidates is required to be greater than twice the vertex resolution away from the IP. The $\Sigma^{0}$ candidates are reconstructed by combining $\Lambda$ candidates with a photon. 
The four-momenta of $K_{S}^{0}$ and $\Lambda$ candidates are updated according to the vertex fits.
Signal mass windows are defined as $487~{\rm MeV}/c^{2} < M(\pi^{+}\pi^{-}) < 511~\rm{MeV}/c^{2}$, $1111~{\rm MeV}/c^{2} < M(p\pi^{-}) < 1121~\rm{MeV}/c^{2}$, and $1179~{\rm MeV}/c^{2} < M(\Lambda \gamma) < 1203~\rm{MeV}/c^{2}$ to select the $K_{S}^{0}$, $\Lambda$, and $\Sigma^{0}$ signal candidates, respectively.

To further improve the $\Lambda^{+}_{c}$ signal resolution, a vertex fit is performed on all charged tracks (including tracks of long-lived particles) to constrain them to one common vertex. Events with a fit quality of $\chi^{2}<200$ are selected as candidate events.

\section{\label{sec:xs_ScSc}Measurements of $e^{+}e^{-}\to \Sigma_{c}\bar{\Sigma}_{c}$ cross section}

\subsection{\label{sec:xs_ScSc_selection}Event selection}

\paragraph{Tag $\Lambda^{+}_{c}$ sample}
Among all the $\Lambda_{c}^{+}$ candidates reconstructed in an event, only the one yielding the minimum $|\Delta M| = |M(\Lambda^{+}_{c}) - m(\Lambda^{+}_{c})|$ value is retained for subsequent analysis, where $M(\Lambda^{+}_{c})$ represents the reconstructed invariant mass of the $\Lambda^{+}_{c}$ candidate and $m(\Lambda^{+}_{c})$ is the known $\Lambda^{+}_{c}$ mass~\cite{cit:PDG_value}. Furthermore, the requirement $|\Delta M| < 0.02~\mathrm{GeV}/c^{2}$ is applied to reject non-$\Lambda^{+}_{c}$ background. The value $RM(\Lambda^{+}_{c}) + M(\Lambda^{+}_{c}) - m(\Lambda^{+}_{c})$, corresponding to the $\Lambda^{+}_{c}$ recoil mass $RM(\Lambda^{+}_{c})$ subtracted by the $\Lambda^{+}_{c}$ mass resolution, must exceed $2.54~\mathrm{GeV}/c^{2}$ to exclude processes like $e^{+}e^{-}\to\Lambda_{c}^{+}\bar{\Lambda}^{-}_{c}$ and $e^{+}e^{-}\to \Lambda_c^{+}\bar{\Sigma}_{c}^{-}$, where $RM(\Lambda^{+}_{c}) = |P_{e^{+}e^{-}} - P_{\Lambda^{+}_{c}}|/c$, and $P_{e^{+}e^{-}}$ and $P_{\Lambda^{+}_{c}}$ represent the four-momenta of $e^{+}e^{-}$ system and $\Lambda^{+}_{c}$ in the laboratory frame, respectively. 

\paragraph{Tag $\Lambda^{+}_{c}\pi$ sample}
The $\Lambda^{+}_{c}$  candidates are combined with residual  $\pi^{+}$, $\pi^{-}$, or $\pi^{0}$ to form $\Sigma_{c}$ candidates.
The $\pi^{0}$ candidates are reconstructed from photon pairs with invariant masses $M(\gamma\gamma)$ ranging from 115~MeV/$c^{2}$ to 150~MeV/$c^{2}$. To improve the momentum resolution, a kinematic fit constraining the invariant mass of the photon pairs to the $\pi^{0}$ known mass~\cite{cit:PDG_value} is performed and the resulting four-momentum of the $\pi^{0}$ candidate is utilized for further analysis.

After applying the $\Lambda^{+}_{c}$ mass window requirement, $M(\Lambda^{+}_{c})\in(2.275,2.300)~\mathrm{GeV}/c^{2}$, the optimal $\Lambda^{+}_{c}\pi$ candidate is selected by minimizing the variable $\Delta E(\Lambda^{+}_{c}\pi) = E_{{\rm CMS}} - E(\Lambda^{+}_{c}\pi) - E_{{\rm rec}}(\Lambda^{+}_{c}\pi)$, where $E_{{\rm CMS}}$ represents the total energy of the initial $e^+e^-$ collisions, $E(\Lambda^{+}_{c}\pi)$ is the total energy of $\Lambda^{+}_{c}$ and $\pi$, and $E_{{\rm rec}}(\Lambda^{+}_{c}\pi) = \sqrt{ p^{2}_{\Lambda^{+}_{c}\pi}c^2 + m(\Sigma_{c})^{2}c^4}$, where $p_{\Lambda^{+}_{c}\pi}$ is the total momentum of $\Lambda^{+}_{c}$ and $\pi$, and $m(\Sigma_{c})$ is the known $\Sigma_{c}$ mass~\cite{cit:PDG_value}. In order to suppress  wrong combinations and hadronic background, $\Delta E(\Lambda^{+}_{c}\pi)$ is required to be greater than $-0.018~\mathrm{GeV}$. This criterion also improves the $\Lambda^{+}_{c}\bar{\Lambda}_{c}^{*-}$ resolution with minimal signal efficiency loss. To take advantage of the accuracy of the beam energy $E_{{\rm beam}}$ measurement~\cite{cit:4750_4946_data_lumin},  the beam-constrained mass $M_{\mathrm{BC}}(\Lambda^{+}_{c}\pi)c^{2} = \sqrt{E_{{\rm beam}}^{2} - p^{2}_{\Lambda^{+}_{c}\pi}c^{2}}$ is used for the subsequent fit.

\paragraph{Tag $\Lambda^{+}_{c}\pi\pi$ sample}
This sample is formed  by combining the residual $\pi^+\pi^-$ or $\pi^0\pi^0$ pairs with a $\Lambda^{+}_{c}$ candidate. The variable $\Delta E(\Lambda^{+}_{c}\pi\pi) =  E_{{\rm CMS}} - E(\Lambda^{+}_{c}\pi\pi) - E_{{\rm rec}}(\Lambda^{+}_{c}\pi\pi)$ is constructed to select the optimal $\Lambda^{+}_{c}\pi\pi$ candidates, where $E(\Lambda^{+}_{c}\pi\pi)$ is the total energy of $\Lambda^{+}_{c}$ and $\pi\pi$, and $E_{{\rm rec}}(\Lambda^{+}_{c}\pi\pi) = \sqrt{ p^{2}_{\Lambda^{+}_{c}\pi\pi}c^2 + m(\Lambda^{+}_{c})^{2}c^4}$. The condition $\Delta E(\Lambda^{+}_{c}\pi\pi)>-0.02~\mathrm{GeV}$ is required for background suppression, and signal candidates with $RM(\Lambda^{+}_{c}\pi\pi) + M(\Lambda^{+}_{c}) - m(\Lambda^{+}_{c})$ within the range $(2.26,2.31)~\mathrm{GeV}/c^{2}$ are retained to identify the $\bar{\Lambda}^{-}_{c}$ on the recoil side.

\subsection{\label{sec:xs_ScSc_simfit}Simultaneous fit}

An unbinned maximum likelihood fit is performed on the distributions across all the tagged samples and energy points. Explicitly, the fit comprises three forms: a 1-D fit to the $RM(\Lambda^{+}_{c}) + M(\Lambda^{+}_{c}) - m(\Lambda^{+}_{c})$ distribution in the ``Tag $\Lambda^{+}_{c}$" sample, a 2-D fit to the $M_{\mathrm{BC}}(\Lambda^{+}_{c}\pi)$ versus $RM(\Lambda^{+}_{c}) + M(\Lambda^{+}_{c}) - m(\Lambda^{+}_{c})$ distribution in the ``Tag $\Lambda^{+}_{c}\pi$" samples, and a 2-D fit to the $M(\Lambda^{+}_{c}\pi\pi) - M(\Lambda^{+}_{c}) + m(\Lambda^{+}_{c})$ versus $RM(\Lambda^{+}_{c}) + M(\Lambda^{+}_{c}) - m(\Lambda^{+}_{c})$ distribution in the ``Tag $\Lambda^{+}_{c}\pi\pi$" samples. The yields of signal processes at each energy point are constrained to their observed cross sections, with the branching fractions of $\Lambda_{c}^{*+}\to\Lambda^{+}_{c}\pi\pi$ as shared parameters across different energy points.

The observed cross sections are determined by
\begin{equation} \label{eq:xs_ScSc_obs_xs}
\sigma_{{\rm obs}} = \frac{N_{{\rm fit}}({\rm Tag~X})}{2\times \mathcal{L}_{{\rm int}}\times \epsilon_{{\rm tot}}(\rm{ Tag~X})},
\end{equation}
where $N_{{\rm fit}}$(Tag X) represents the fitted yield in the ``Tag $\Lambda^{+}_{c} $" and ``Tag $\Lambda^{+}_{c}\pi$" samples, and $\mathcal{L}_{{\rm int}}$ denotes the integrated luminosity obtained from Ref.~\cite{cit:4750_4946_data_lumin}; $\epsilon_{{\rm tot}}({\rm Tag~X}) = \sum_{i=1}^{4} \epsilon_{i}({\rm Tag~X})\times {\cal B}_{i}$ is the sum of the products of the detection efficiency $\epsilon_{i}(\text{Tag X})$ in the ``Tag X" sample and the corresponding branching fraction ${\cal B}_{i}$ for the $i$-th $\Lambda^{+}_{c}$ tag mode~\cite{cit:PDG_value}. Notably, the detection efficiencies include the branching fractions of the $\Lambda^{+}_{c}$ daughter particles, and represent the average efficiencies of $\Lambda_{c}^{+}$ and $\bar{\Lambda}^{-}_{c}$.

Considering the ratio between the yields of $\Lambda_{c}^{*+}$ in ``Tag $\Lambda^{+}_{c}\pi\pi$" samples $N^{\Lambda_{c}^{*+}}_{{\rm fit}}({\rm Tag}~\Lambda^{+}_{c}\pi\pi)$ and ``Tag $\Lambda^{+}_{c}$" sample $N^{\Lambda_{c}^{*+}}_{{\rm fit}}({\rm Tag}~\Lambda^{+})$, along with the efficiency, the inclusive branching fractions of $\Lambda_{c}^{*+}\to\Lambda_{c}^{+}\pi^{+}\pi^{-}$ and $\Lambda_{c}^{*+}\to\Lambda_{c}^{+}\pi^{0}\pi^{0}$ can be formulated as follows
\begin{equation}\label{eq:xs_ScSc_br_LcPiPi}
{\cal B}(\Lambda_{c}^{*+}\to\Lambda_{c}^{+}\pi\pi)=\frac{{\cal A}^{\Lambda_{c}^{*+}}({\rm Tag}~\Lambda^{+}_{c}\pi\pi)}{{\cal A}^{\Lambda_{c}^{*+}}({\rm Tag}~\Lambda^{+}_{c})},
\end{equation}
where ${\cal A}^{\Lambda_{c}^{*+}}=N^{\Lambda_{c}^{*+}}_{{\rm fit}}/ \epsilon^{\Lambda_{c}^{*+}}_{{\rm tot}}$.

Since the use of simultaneous fit is not able to constrain the large number of free parameters with the current statistics, the isospin symmetry of the following cross sections are assumed, i.e., $\sigma_{{\rm obs}}(e^{+}e^{-}\to\Sigma_{c}^{+}\bar{\Sigma}_{c}^{-})=\sigma_{{\rm obs}}(e^{+}e^{-}\to\Sigma_{c}^{0}\bar{\Sigma}_{c}^{0})=\sigma_{{\rm obs}}(e^{+}e^{-}\to\Sigma_{c}^{++}\bar{\Sigma}_{c}^{--})$, and $\sigma_{{\rm obs}}(e^{+}e^{-}\to\Lambda^{+}_{\it c}\bar{\Sigma}^{0}_{\it c}\pi^{-}) = \sigma_{{\rm obs}}(e^{+}e^{-}\to\Lambda^{+}_{\it c}\bar{\Sigma}^{--}_{\it c}\pi^{+}) = \sigma_{{\rm obs}}(e^{+}e^{-}\to\Lambda^{+}_{\it c}\bar{\Sigma}^{-}_{\it c}\pi^{0})$.  In addition, according to Refs.~\cite{cit:PDG_value,cit:Belle_LcLcST_br}, the branching fraction ratios of the subprocesses $\Lambda_{c}^{*}(2595)^+(\Lambda_{c}^{*}(2625)^+)\to\Sigma_{c}^{0}\pi^{+}$, $\Sigma_{c}^{++}\pi^{-}$, and $\Lambda_{c}^{+}\pi^{+}\pi^{-}$ (three-body decay) are assumed to be 0.36(0.08), 0.36(0.08), and 0.28(0.84), respectively, to the inclusive branching fraction of $\Lambda_{c}^{*+}\to\Lambda_{c}^{+}\pi^{+}\pi^{-}$. Similarly, for the final state $\Lambda_{c}^{+}\pi^{0}\pi^{0}$, the ratios are assumed to be 0.71(0.15) for $\Lambda_{c}^{*+}\to\Sigma_{c}^{+}\pi^{0}$ and 0.29(0.85) for $\Lambda_{c}^{+}\pi^{0}\pi^{0}$ (three-body decay). The constraints will be factored into the efficiency estimation and the construction of the signal model. 

The signal shapes are obtained from MC simulations convolved with Gaussian functions, to take into account the resolution discrepancy between data and MC simulations. The parameters of the Gaussian functions on different dimensions are fixed according to studies with control samples.  In the $RM(\Lambda^{+}_{c}) + M(\Lambda^{+}_{c}) - m(\Lambda^{+}_{c})$ distribution, the data control samples $e^{+}e^{-}\to\Lambda_{c}^{+}\bar{\Lambda}_{c}^{-}$ at $\sqrt{s} = 4.600$, $4.612$, and $4.641~\mathrm{GeV}$ are used to assess the resolution discrepancy in the signal process, based on the tagged $\Lambda^{+}_{c}$ momentum.
The same method is used to estimate the resolution discrepancy in the $M_{\mathrm{BC}}(\Lambda^{+}_{c}\pi)$ distributions.
The resolution discrepancy in the $M(\Lambda^{+}_{c}\pi\pi) - M(\Lambda^{+}_{c}) + m(\Lambda^{+}_{c})$ distributions is evaluated by analyzing the decay $\psi(3686)\to\pi\pi J/\psi$, $J/\psi\to\rho\pi$. Gaussian functions in the $M(J/\psi \pi\pi) - M(J/\psi) + m(J/\psi)$ spectra are employed to characterize the resolution effect of the signal process, with $M(J/\psi \pi\pi)$, $M(J/\psi)$, and $m(J/\psi)$ representing the invariant mass of $J/\psi \pi\pi$ and $J/\psi$, and the known mass of $J/\psi$~\cite{cit:PDG_value}, respectively.

The backgrounds from $\Lambda_c^+$ pair productions and non-$\Lambda_c^+$ hadronic backgrounds are evaluated according to the MC simulations. Their shapes are obtained from the corresponding MC samples. In the fit, the amplitudes of the $\Lambda_c^+$ pair backgrounds are fixed according to the BESIII measurement~\cite{cit:Bes_LcLc_xs_new}, while the contributions of hadronic backgrounds are scaled from MC simulations by a global factor across all tagged samples at a single energy point, which is left floating in the fit. 

\begin{figure}[htbp]
\centering
\begin{minipage}{0.35\textwidth}
  \centering
  \includegraphics[width=6.2cm, height=4.6cm]{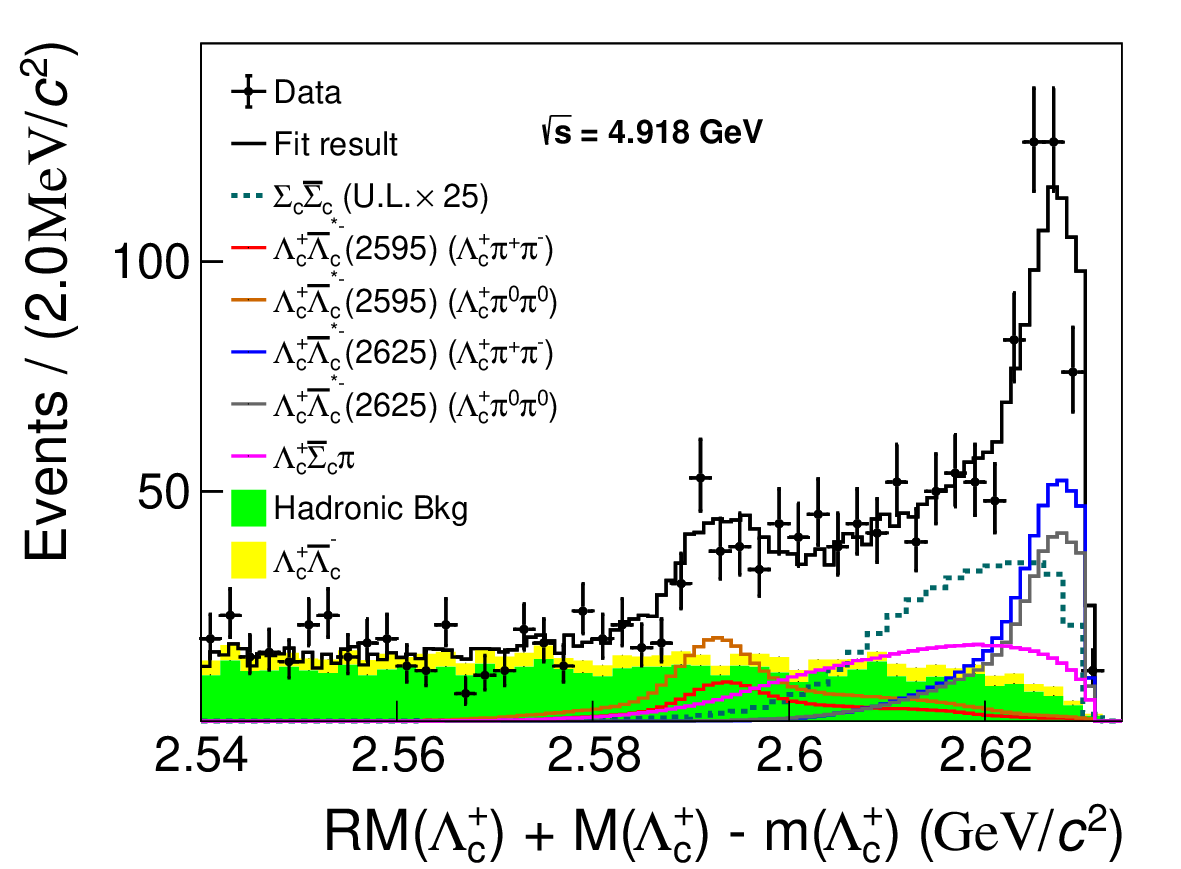}
\end{minipage}%

\begin{minipage}{0.35\textwidth}
  \centering
  \includegraphics[width=6.2cm, height=4.6cm]{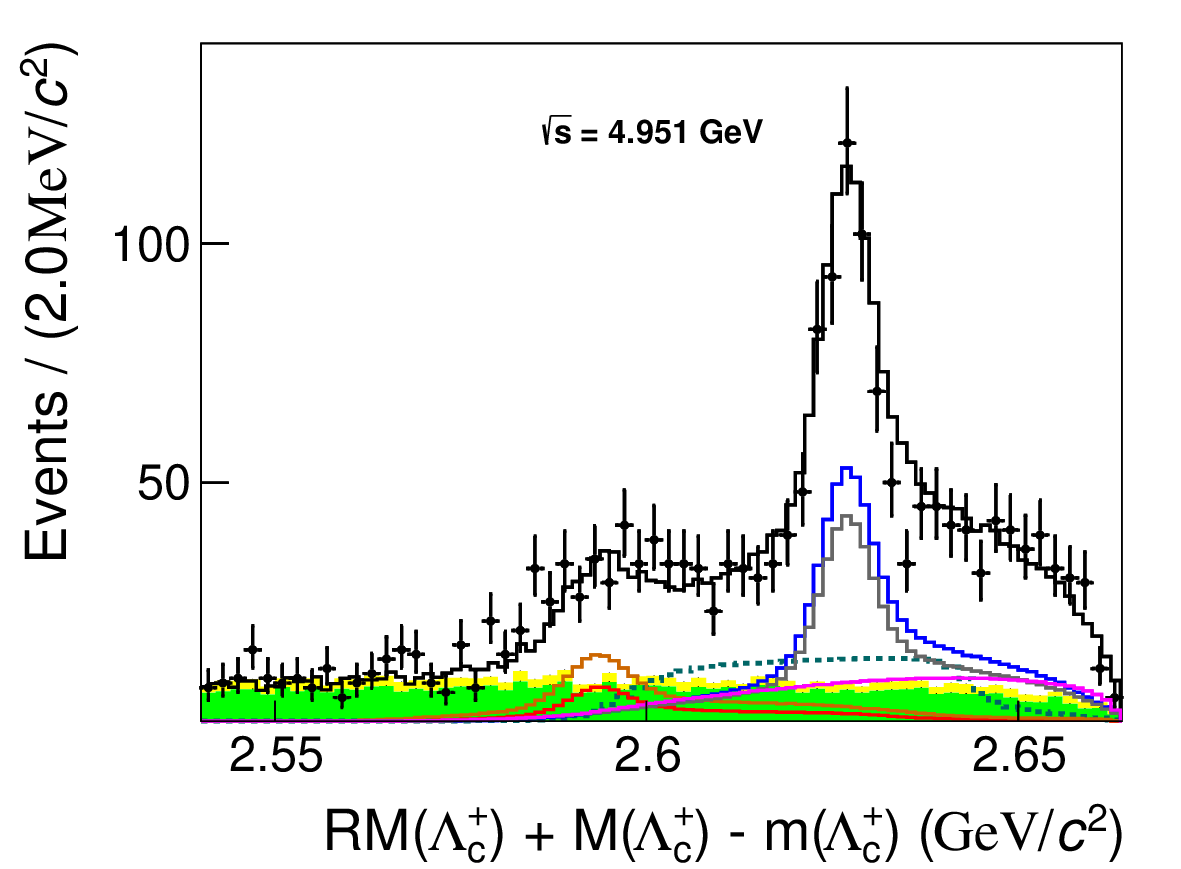}
\end{minipage}
\caption{One-dimensional fit results at $\sqrt{s} = 4.918~\mathrm{GeV}$ (top) and $\sqrt{s} = 4.951~\mathrm{GeV}$ (bottom) in the ``Tag $\Lambda^{+}_{c}$" sample. The dots with error bars denote data,  and the black solid curves correspond to the fit result.  ``Bkg" stays for the background. The dashed lines represent the  MC distributions of the components $\Sigma_{c}^{+}\bar{\Sigma}_{c}^{-}, \Sigma_{c}^{0}\bar{\Sigma}_{c}^{0}$ and $\Sigma_{c}^{++}\bar{\Sigma}_{c}^{--}$. The ``U.L. $\times$ 25" indicates the MC distribution normalized to 25 times  the 90\% confidence level upper limit of the $\Sigma_{c}\bar{\Sigma}_{c}$ observed cross section.
}
\label{fig:fit_tagLc}
\end{figure}

\begin{figure*}[htbp]
\centering
\begin{minipage}{0.48\textwidth}
  \centering
  \includegraphics[width=8.6cm, height=3.5cm]{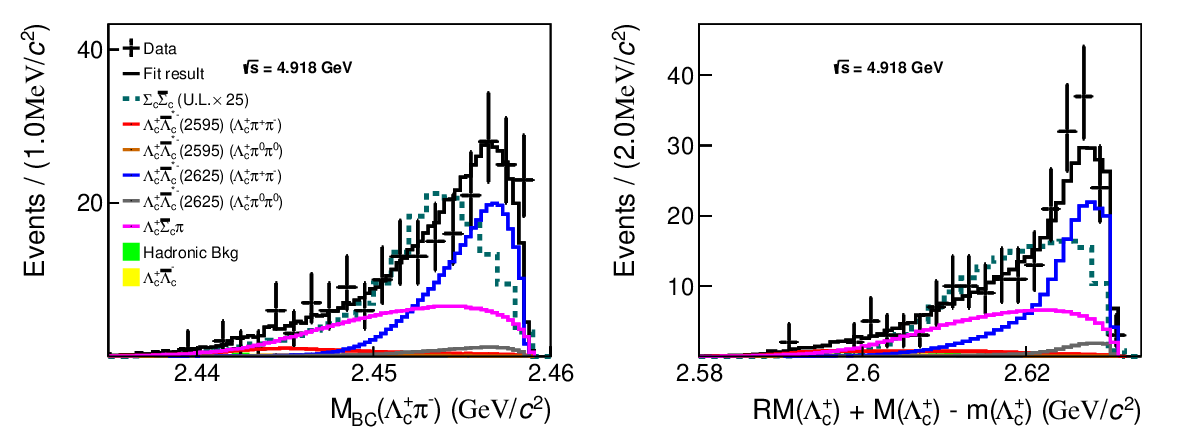}
\end{minipage}%
\hspace{0.001\textwidth}
\begin{minipage}{0.48\textwidth}
  \centering
  \includegraphics[width=8.6cm, height=3.5cm]{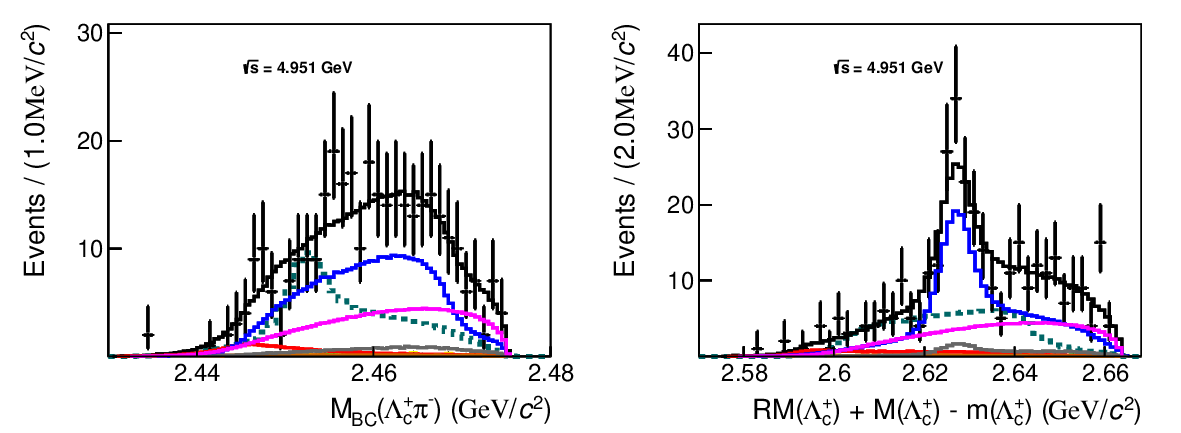}
\end{minipage}
\begin{minipage}{0.48\textwidth}
  \centering
  \includegraphics[width=8.6cm, height=3.5cm]{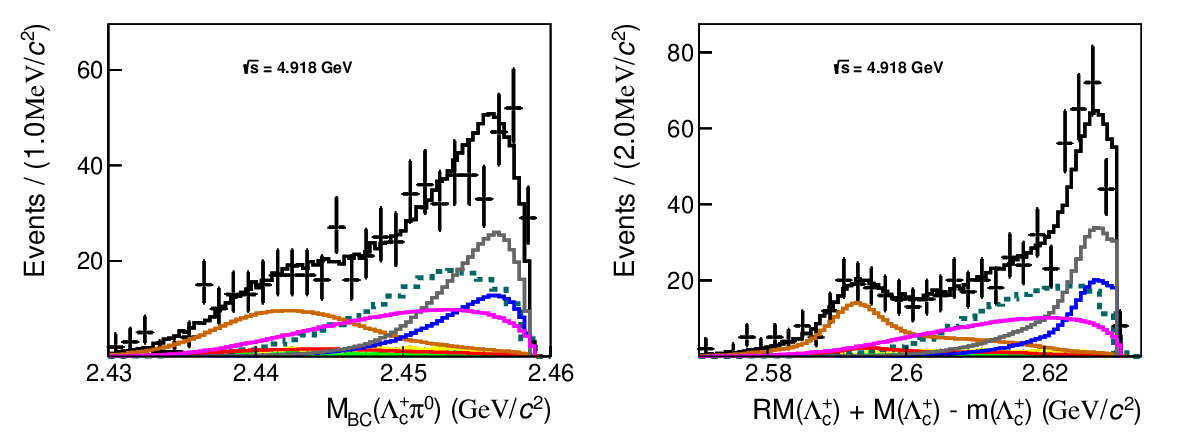}
\end{minipage}%
\hspace{0.001\textwidth}
\begin{minipage}{0.48\textwidth}
  \centering
  \includegraphics[width=8.6cm, height=3.5cm]{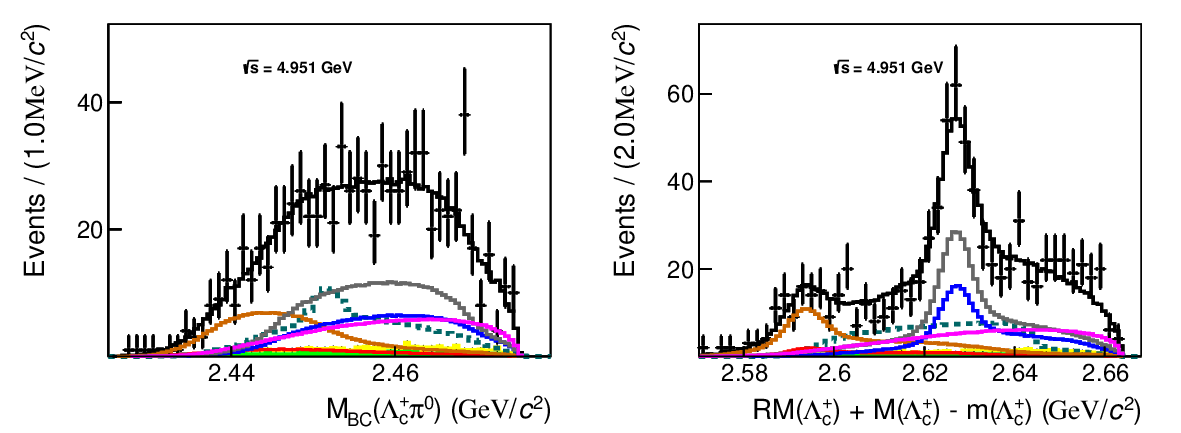}
\end{minipage}
\begin{minipage}{0.48\textwidth}
  \centering
  \includegraphics[width=8.6cm, height=3.5cm]{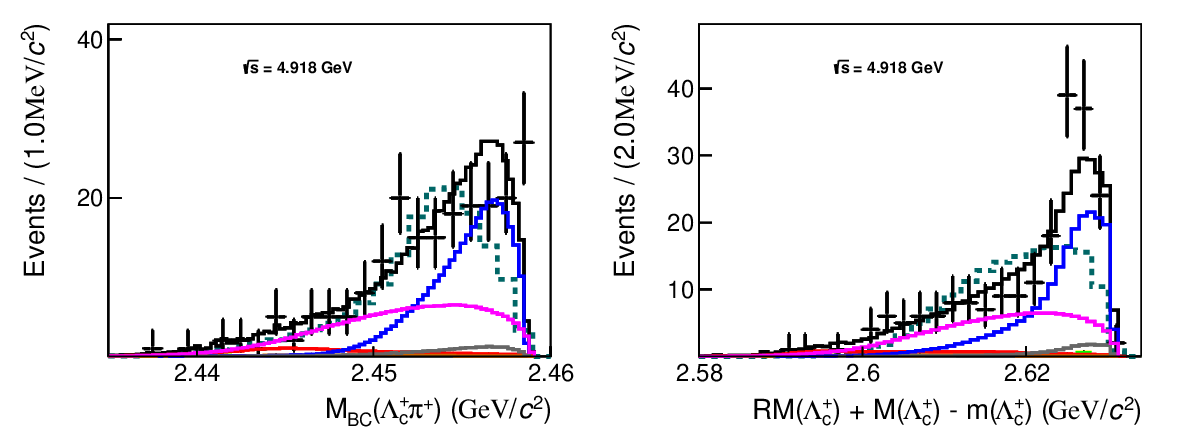}
\end{minipage}%
\hspace{0.001\textwidth}
\begin{minipage}{0.48\textwidth}
  \centering
  \includegraphics[width=8.6cm, height=3.5cm]{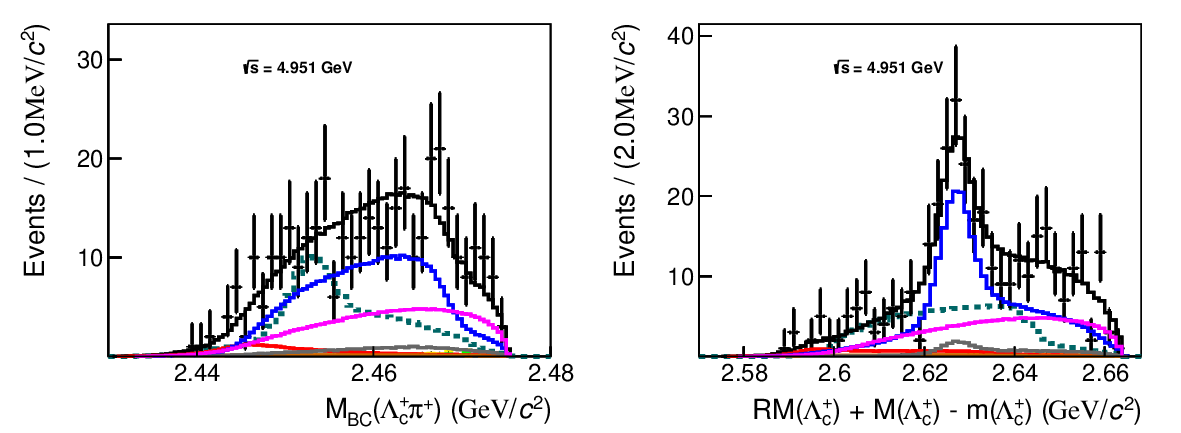}
\end{minipage}
\begin{minipage}{0.48\textwidth}
  \centering
  \includegraphics[width=8.6cm, height=3.5cm]{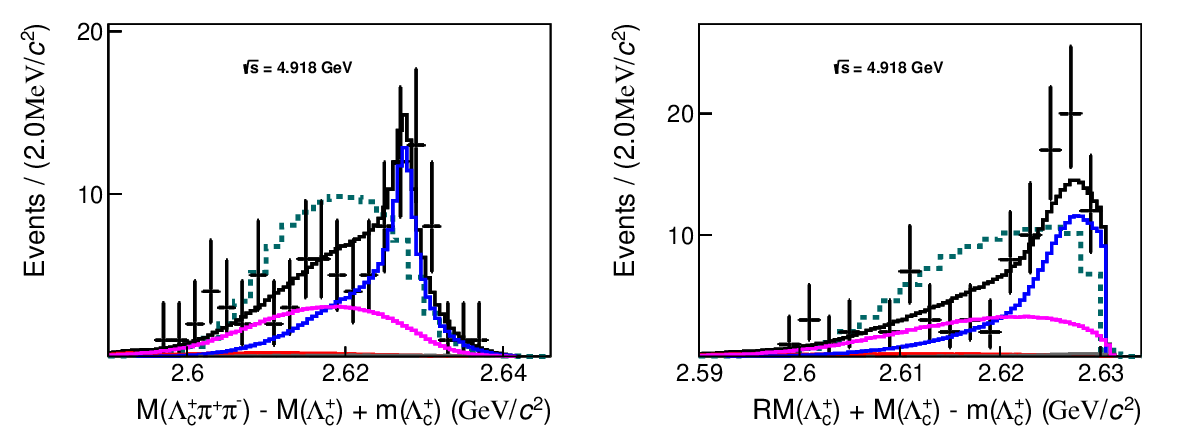}
\end{minipage}%
\hspace{0.001\textwidth}
\begin{minipage}{0.48\textwidth}
  \centering
  \includegraphics[width=8.6cm, height=3.5cm]{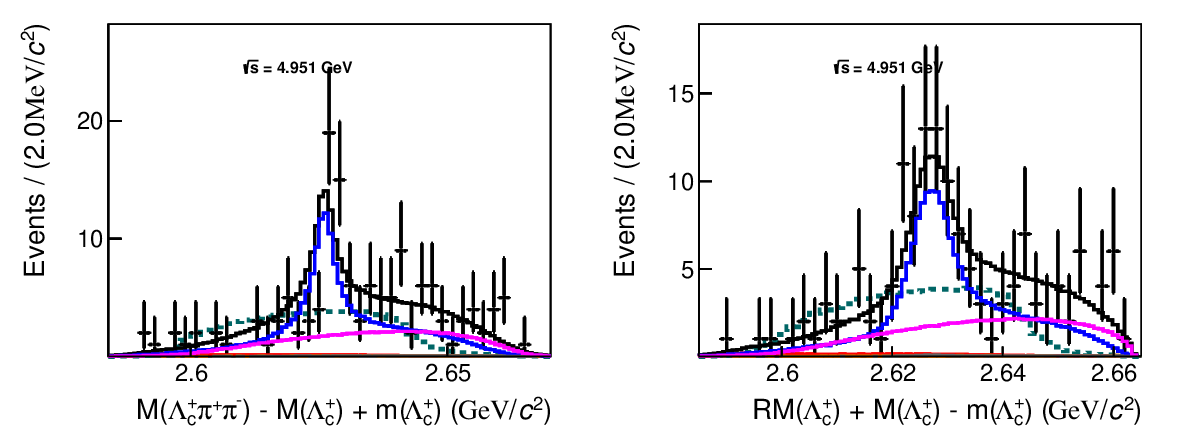}
\end{minipage}
\begin{minipage}{0.48\textwidth}
  \centering
  \includegraphics[width=8.6cm, height=3.5cm]{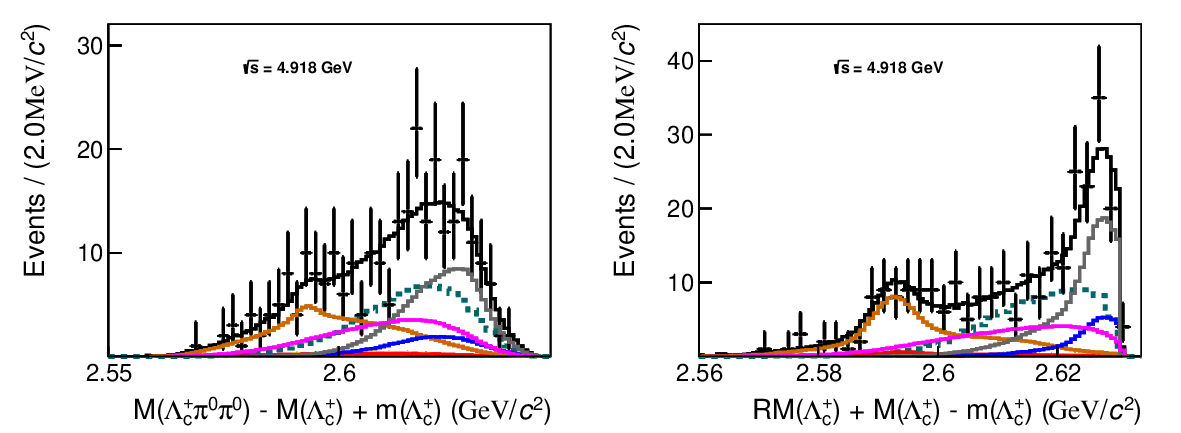}
\end{minipage}%
\hspace{0.001\textwidth}
\begin{minipage}{0.48\textwidth}
  \centering
  \includegraphics[width=8.6cm, height=3.5cm]{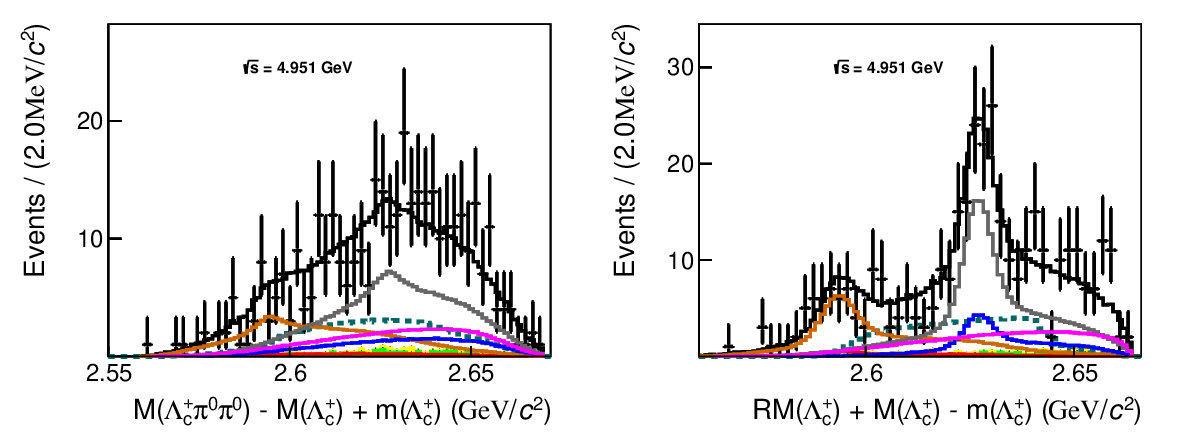}
\end{minipage}
\caption{Two-dimensional fit results at $\sqrt{s} = 4.918$ and $4.951~\mathrm{GeV}$. The rows, from top to bottom, correspond to the results in the samples of ``Tag $\Lambda^{+}_{c}\pi^{-}$", ``Tag $\Lambda^{+}_{c}\pi^{0}$", ``Tag $\Lambda^{+}_{c}\pi^{+}$", ``Tag $\Lambda^{+}_{c}\pi^{+}\pi^{-}$", and ``Tag $\Lambda^{+}_{c}\pi^{0}\pi^{0}$". The dots with error bars denote the data, while the black solid curves correspond to the total fit function. ``Bkg" stays for the background. The dashed lines represent the MC distributions of the components $\Sigma_{c}^{+}\bar{\Sigma}_{c}^{-}, \Sigma_{c}^{0}\bar{\Sigma}_{c}^{0}$ and $\Sigma_{c}^{++}\bar{\Sigma}_{c}^{--}$. The ``U.L. $\times$ 25" indicates  the MC distributions  normalized to 25 times  the 90\% confidence level upper limit of the $\Sigma_{c}\bar{\Sigma}_{c}$ observed cross section.}
\label{fig:fit_tagLcPi_tagLcPiPi}
\end{figure*}

The fit results are illustrated in Fig.~\ref{fig:fit_tagLc} and Fig.~\ref{fig:fit_tagLcPi_tagLcPiPi}, where no $e^{+}e^{-}\to\Sigma_{c}\bar{\Sigma}_{c}$ signal is present at either $\sqrt{s} = 4.918~\mathrm{GeV}$ or $\sqrt{s} = 4.951~\mathrm{GeV}$. 
Hence, the upper limits on the observed cross section $\sigma_{{\rm U.L.}}$ at the 90$\%$ confidence level (C.L.) are obtained by solving the equation:
\begin{equation}\label{eq:UL}
\int_{0}^{ \sigma_{{\rm U.L.}} } \mathcal{L}(x) \mathrm{d}x = 0.9 \int_{0}^{ \infty } \mathcal{L}(x) \mathrm{d}x,
\end{equation}
where $x$ is the assumed cross section of $e^{+}e^{-}\to\Sigma_{c}\bar{\Sigma}_{c}$ and $\mathcal{L}(x)$ is the maximized likelihood of the data assuming a cross section of $x$. 

The statistical correlations among different data samples are evaluated through tests with the simulated MC samples. The input-output (IO) checks reveal that the current fit strategy underestimates the uncertainty and introduces biases in the central values. However, this underestimation has a negligible effect on the upper limit estimation. As a result, only the bias on $x$ in Eq.~\ref{eq:UL} is considered.

The Born cross section is determined by
\begin{equation} \label{eq:Born_xs}
\sigma_{\rm Born} = \frac{\sigma_{{\rm U.L.}}}{f_{\rm VP}\times f_{\rm ISR}},
\end{equation}
where the vacuum polarization (VP) correction factor $f_{{\rm VP}}$ is calculated to be 1.06 at all energy points~\cite{cit:f_VP}. The ISR correction factor $f_{\rm ISR}$~\cite{cit:f_ISR} is obtained from the {\sc KKMC}  generator and calculated as 0.96 at all energy points, assuming the input Born cross section is the same as that of $e^{+}e^{-}\to\Lambda_{c}^{+}\bar{\Lambda}_{c}^{-}$~\cite{cit:Bes_LcLc_xs_new}.

\begin{figure}[htbp]
\includegraphics[width=6.2cm, height=4.6cm]{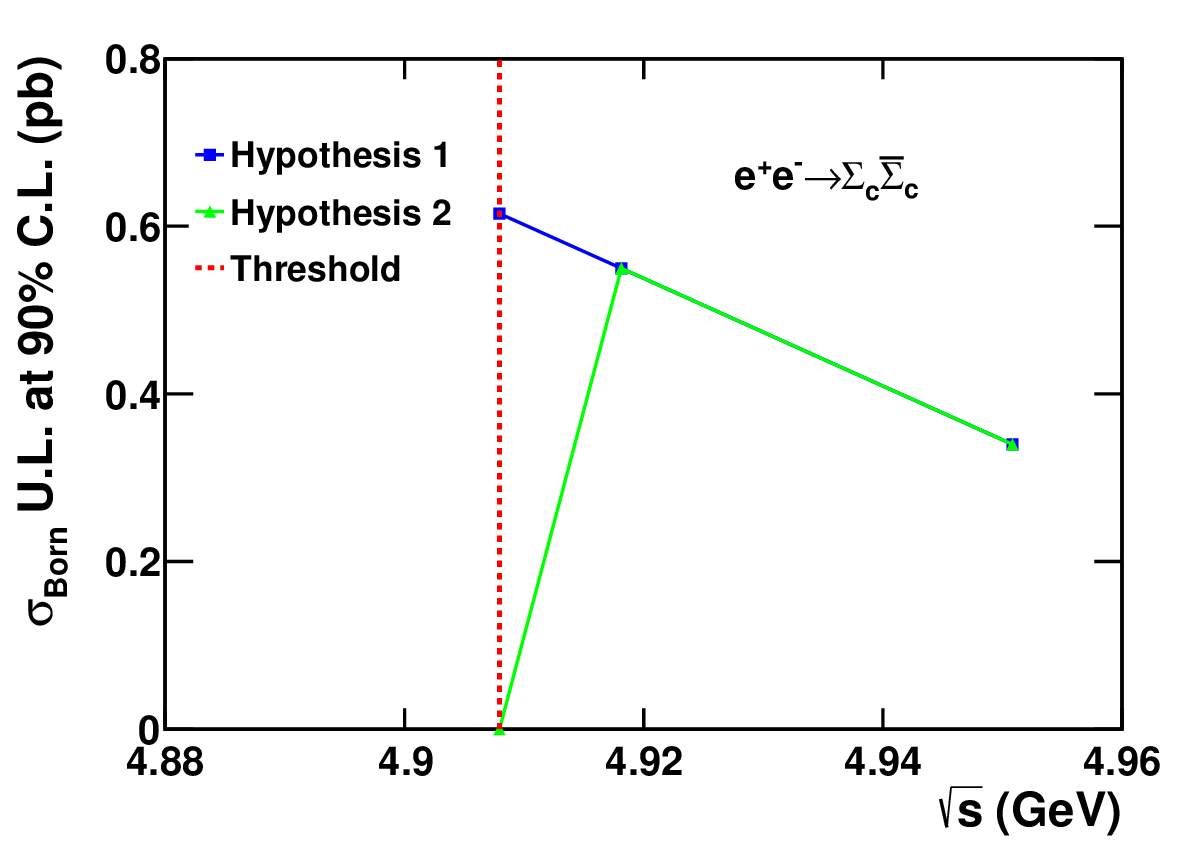}
\caption{\label{fig:xs_ScSc_Born_xs_Line}Upper limit at the 90\% C.L. on Born cross section of $e^{+}e^{-}\to\Sigma_{c}\bar{\Sigma}_{c}$. Points above the threshold represent the upper limit results based on the assumed line shape of the $e^{+}e^{-}\to\Lambda_{c}^{+}\bar{\Lambda}_{c}^{-}$, as shown in Table~\ref{tab:ScSc_Born_xs_different_lineshape}. The blue and green curves represent alternative assumed line shapes.}
\end{figure}

\subsection{\label{sec:xs_ScSc_sys_err}Systematic uncertainty}

The systematic uncertainties can be categorized into multiplicative and additive forms. Details are discussed in the following parts.

\subsubsection{Multiplicative items}

The sources of multiplicative uncertainties arise from $\Lambda^{+}_{c}$ reconstruction, branching fractions of  the $\Lambda^{+}_{c}$ decay channels, integrated luminosity, tracking and PID of residual charged pions, reconstruction of residual $\pi^{0}$, MC statistics, $f_{\rm ISR}$, and $f_{\rm VP}$. Specifically, the uncertainties associated with the reconstruction of $\Lambda_{c}^{+}\to p K_{S}^{0}$, $pK^{-}\pi^{+}$, $\Lambda\pi^{+}$, and $\Sigma^{0}\pi^{+}$, as detailed in Section~\ref{sec:pre_selection}, are estimated as $2.0\%$, $3.9\%$, $2.4\%$, and $2.4\%$, respectively~\cite{cit:Lc_rec_err}, while the uncertainties of branching fractions are assigned as $5.0\%$, $5.0\%$, $5.3\%$, and $5.4\%$,  respectively~\cite{cit:PDG_value}. According to Ref.~\cite{cit:4750_4946_data_lumin}, the uncertainties associated with the integrated luminosity are estimated as $0.5\%$ for all energy points. The uncertainties in tracking and PID for residual charged pions are primarily dependent on the pions' momenta. Discrepancies in their efficiencies between data and MC simulation are determined to be $3.5\%$ and $1.6\%$, respectively, from the analysis of pions with momenta below 0.2~${\rm GeV}/c$ in the hadronic decays of $D^{0}$ and $D^{\pm}$ mesons. For the reconstruction of residual $\pi^{0}$, the uncertainty is assigned as $2.8\%$, according to the study of $\pi^{0}$ with momenta below 0.2 GeV in the control channel $D^{0}\to K^{-}\pi^{+}\pi^{0}$. The uncertainty associated with MC statistics is $0.2\%$ for all processes. 

Several sources of systematic uncertainties are considered in the determination of $f_{\rm ISR}$. Firstly, the uncertainties arising from the model choice are estimated by employing  {\sc ConExc}~\cite{cit:ConExc} as alternative algorithm. They are determined to be 6.9\% and 0.1\% at $\sqrt{s}$ = 4.918 and 4.951 GeV, respectively. Subsequently, the uncertainty associated with the data energy~\cite{cit:4750_4946_data_lumin} is investigated by varying the input energy within its error range in the MC simulation. This uncertainty is quantified as 1.1\% and 0.3\% at $\sqrt{s}$ = 4.918 and 4.951 GeV, respectively. Similarly, the systematic uncertainty arising from the estimation of the beam energy spread~\cite{cit:beam_spread} is determined to be 1.5\% and 0.2\% at $\sqrt{s}$ = 4.918 and 4.951 GeV, respectively. 
The uncertainty related to the calculation of $f_{\rm VP}$ at all energy points is determined as 0.5$\%$~\cite{cit:f_VP}.

A summary of multiplicative uncertainties except for $f_{\rm ISR}$ and $f_{\rm VP}$ is presented in Table~\ref{tab:sys_multi_err}, and they are the same at both energy points.
A MC approach is employed to estimate the resultant effect on the fitted cross sections introduced by the systematic uncertainties, assuming a Gaussian distribution for each multiplicative uncertainty. The total multiplicative uncertainties are determined to be 13.1\% and 8.1\% for the Born cross section at $\sqrt{s}=4.918$ and 4.951 GeV, respectively.
\begin{table*}[htbp]
\centering
\caption{ \label{tab:sys_multi_err} Summary of the multiplicative uncertainties (in \%), except for $f_{\rm ISR}$ and $f_{\rm VP}$. ``Rec" is an abbreviation for reconstruction, while $\cal{B}$ denotes the branching fraction. ``N/A" means this term is not applicable to the corresponding sample.}
\begin{ruledtabular}
\begin{tabular}{lccccccccc}
\multirow{2}{*}{Sample} & \multirow{2}{*}{$\mathcal{L}_{{\rm int}}$} & \multicolumn{4}{c}{$\Lambda^{+}_{c}$ Rec~($\cal{B}$)}  
& \multirow{2}{*}{$\pi^{+}$ Tracking} & \multirow{2}{*}{$\pi^{+}$ PID} & \multirow{2}{*}{$\pi^{0}$ Rec} &\multirow{2}{*}{MC Statistics} \\
\cmidrule{3-6}
~&~& \small{$p K_{S}^{0}$} & \small{$pK^{-}\pi^{+}$} & \small{$\Lambda\pi^{+}$} & \small{$\Sigma^{0}\pi^{+}$} &~&~&~&~\\
Tag $\Lambda^{+}_{c}$ & $0.5$ &$2.0~(5.0)$&$3.9~(5.0)$&$2.4~(5.3)$&$2.4~(5.4)$&N/A&N/A&N/A&$0.2$\\
Tag $\Lambda^{+}_{c}\pi^{\pm}$ & $0.5$ &$2.0~(5.0)$&$3.9~(5.0)$&$2.4~(5.3)$&$2.4~(5.4)$&$3.5$&$1.6$&N/A&$0.2$\\
Tag $\Lambda^{+}_{c}\pi^{0}$ & $0.5$ &$2.0~(5.0)$&$3.9~(5.0)$&$2.4~(5.3)$&$2.4~(5.4)$&N/A&N/A&$2.8$&$0.2$\\
Tag $\Lambda^{+}_{c}\pi^{+}\pi^{-}$ & $0.5$ &$2.0~(5.0)$&$3.9~(5.0)$&$2.4~(5.3)$&$2.4~(5.4)$&$7.0$&$3.2$&N/A&$0.2$\\
Tag $\Lambda^{+}_{c}\pi^{0}\pi^{0}$ & $0.5$ &$2.0~(5.0)$&$3.9~(5.0)$&$2.4~(5.3)$&$2.4~(5.4)$&N/A&N/A&$5.6$&$0.2$\\
\end{tabular}
\end{ruledtabular}
\end{table*}

\subsubsection{Additive items}

The additive uncertainties consist of the line shape of the Born cross section for $e^{+}e^{-}\to\Sigma_{c}\bar{\Sigma}_{c}$, the polarization of $e^{+}e^{-}\to\Lambda_{c}^{+}\bar{\Lambda}_{c}^{*-}$, the polarization of $e^{+}e^{-}\to\Sigma_{c}\bar{\Sigma}_{c}$, isospin constraints among the cross sections of $e^{+}e^{-}\to\Lambda_{c}\bar{\Sigma}_{c}\pi$ and among the cross sections of $e^{+}e^{-}\to\Sigma_{c}\bar{\Sigma}_{c}$, and the correction for the difference in resolution between the simulated and real data samples.  

The uncertainties in the cross section line shape arise from variations in the ISR correction. Given the absence of any observed signal and the lack of theoretical predictions, the baseline ISR correction factor (summarized in Table~\ref{tab:ScSc_Born_xs_different_lineshape}) is computed using the measured $e^{+}e^{-}\to\Lambda_{c}^{+}\bar{\Lambda}_{c}^{-}$ cross section from Ref.~\cite{cit:Bes_LcLc_xs_new}. Systematic uncertainties associated with the $e^{+}e^{-}\to\Sigma_{c}\bar{\Sigma}_{c}$ line shape are subsequently derived by analyzing line shapes under two new hypotheses. Hypothesis 1 holds that the production of $e^{+}e^{-}\to\Sigma_{c}\bar{\Sigma}_{c}$ exhibits a threshold enhancement, similar to cases such as $e^{+}e^{-}\to p\bar{p}$~\cite{cit:xs_exp_3_pp_xs} and $e^{+}e^{-}\to\Lambda_{c}^{+}\bar{\Lambda}_{c}^{-}$~\cite{cit:Bes_LcLc_xs_new}. Hypothesis 2 assumes no threshold enhancement effect, and considers the most extreme scenario where the cross section is zero at the threshold. The line shapes that correspond to both hypotheses are represented in Fig.~\ref{fig:xs_ScSc_Born_xs_Line}.
The Born cross sections based on each hypothesis are presented in Table~\ref{tab:ScSc_Born_xs_different_lineshape}. The comparison reveals that, under the Hypothesis 2 in Table~\ref{tab:ScSc_Born_xs_different_lineshape}, the upper limit estimation of the $e^{+}e^{-}\to\Sigma_{c}\bar{\Sigma}_{c}$ Born cross section reaches its most conservative value. Thus, subsequent studies on additive systematic uncertainties are based on this scenario.

The potential impact of $\Lambda_{c}^{*+}$ polarization is considered, as it can affect the detection efficiency and signal shape of $e^{+}e^{-}\to\Lambda_{c}^{+}\bar{\Lambda}_{c}^{*-}$,  ultimately influencing the evaluation of $e^{+}e^{-}\to\Sigma_{c}\bar{\Sigma}_{c}$ signal yields. To evaluate the potential effect, the events in the signal phase space MC samples are weighted based on the angular distribution form
$f(\theta) \propto 1 + \alpha_{\Lambda^{+}_{c}}\cos^{2}\theta_{\Lambda^{+}_{c}}$~\cite{cit:Bes_LcLc_xs_old}.
Here, $\theta_{\Lambda^{+}_{c}}$ is the polar angle of $\Lambda^{+}_{c}$ in the  CM system of its mother particle $\Lambda_{c}^{*+}$. The parameter $\alpha_{\Lambda^{+}_{c}}$ quantifies the degree of polarization, obtained from Ref.~\cite{cit:Lc2625_polar}. 
In the same manner, the polarization effect of $e^{+}e^{-}\to\Sigma_{c}\bar{\Sigma}_{c}$ can be determined. However, there is no theoretical prediction or experimental observation as inputs. Hence, the extreme scenario is considered, with the $\Sigma_{c}$ being completely polarized at the two energy points, namely $\alpha_{\Sigma_{c}}=\pm 1$.
 
To account for potential isospin symmetry breaking effects, the ratios of $\sigma(e^{+}e^{-}\to\Lambda^{+}_{c}\bar{\Sigma}^{0}_{c}\pi^{-})$,  $\sigma(e^{+}e^{-}\to\Lambda^{+}_{c}\bar{\Sigma}^{--}_{c}\pi^{+})$, and $\sigma(e^{+}e^{-}\to\Lambda^{+}_{c}\bar{\Sigma}^{-}_{c}\pi^{0})$ are varied to two different cases: 1:1:0.5 and 1:1:2, and thereafter the upper limits are re-estimated. Likewise, the ratios of $\sigma(e^{+}e^{-}\to\Sigma_{c}^{+}\bar{\Sigma}_{c}^{-})$, $\sigma(e^{+}e^{-}\to\Sigma_{c}^{++}\bar{\Sigma}_{c}^{--})$, and $\sigma(e^{+}e^{-}\to\Sigma_{c}^{0}\bar{\Sigma}_{c}^{0})$ are changed to 1:1:0.5 and 1:1:2.

The uncertainty due to resolution differences between simulation and data can affect the analysis. To evaluate the potential size, control samples of $e^{+}e^{-}\to\Lambda_{c}^{+}\bar{\Lambda}_{c}^{-}$ at $\sqrt{s}=4.918$ and $4.951$~GeV are used to obtain the new parameters of Gaussian functions for the $RM(\Lambda^{+}_{c}) + M(\Lambda^{+}_{c}) - m(\Lambda^{+}_{c})$ and $M_{\mathrm{BC}}(\Lambda^{+}_{c}\pi)$ distributions. The relevant uncertainties for the signal shapes in the $M(\Lambda^{+}_{c}\pi\pi) - M(\Lambda^{+}_{c}) + m(\Lambda^{+}_{c})$ distributions are tested by removing the smearing Gaussian functions.

Among the above systematic variations, the most conservative UL's, i.e., the cases of $\sigma(e^{+}e^{-}\to\Lambda^{+}_{c}\bar{\Sigma}^{0}_{c}\pi^{-})$,  $\sigma(e^{+}e^{-}\to\Lambda^{+}_{c}\bar{\Sigma}^{--}_{c}\pi^{+})$, and $\sigma(e^{+}e^{-}\to\Lambda^{+}_{c}\bar{\Sigma}^{-}_{c}\pi^{0})$ being in a ratio of 1:1:2, are selected as the final results.

\begin{table*}[htbp]
\centering
\caption{ \label{tab:ScSc_Born_xs_different_lineshape} Summary of $f_{\rm VP}$, $f_{\rm ISR}$ and $\sigma_{\rm Born}$($e^{+}e^{-}\to\Sigma_{c}\bar{\Sigma}_{c}$), based on different assumptions of line shapes: 1. baseline model adopting $e^{+}e^{-}\to\Lambda_{c}^{+}\bar{\Lambda}_{c}^{-}$ measurements from Ref.~\cite{cit:Bes_LcLc_xs_new}; 2. threshold-enhanced hypothesis (Hypothesis 1); 3. non-enhanced scenario (Hypothesis 2). All upper limits are set at the 90\% C.L., and do not include the systematic uncertainties.}
\begin{ruledtabular}
\begin{tabular}{ccccc}
$\sqrt{s}$ & $f$ & baseline & Hypothesis 1 & Hypothesis 2 \\					  
\midrule
\multirow{3}{*}{4.918 GeV} & $f_{\rm VP}$   & 1.06 & 1.06 & 1.06 \\
                                                   &  $f_{\rm ISR}$ & 0.96 & 0.68 & 0.58 \\
                                                   &  $\sigma_{\rm Born}$ & $<$ 0.55 $\rm{pb}$ & $<$ 0.61 $\rm{pb}$ & $<$ 0.83 $\rm{pb}$ \\
\midrule
\multirow{3}{*}{4.951 GeV} & $f_{\rm VP}$   & 1.06 & 1.06 & 1.06\\
                                                   &  $f_{\rm ISR}$ & 0.96 & 0.81 & 0.79 \\
                                                   &  $\sigma_{\rm Born}$ & $<$ 0.34 $\rm{pb}$ & $<$ 0.39 $\rm{pb}$ & $<$ 0.49 $\rm{pb}$ \\
\end{tabular}
\end{ruledtabular}
\end{table*}

\subsection{\label{sec:xs_ScSc_final_result}Final results}

\begin{figure}[htbp]
\centering
\begin{minipage}{0.45\textwidth}
  \centering
  \includegraphics[width=6.2cm, height=4.6cm]{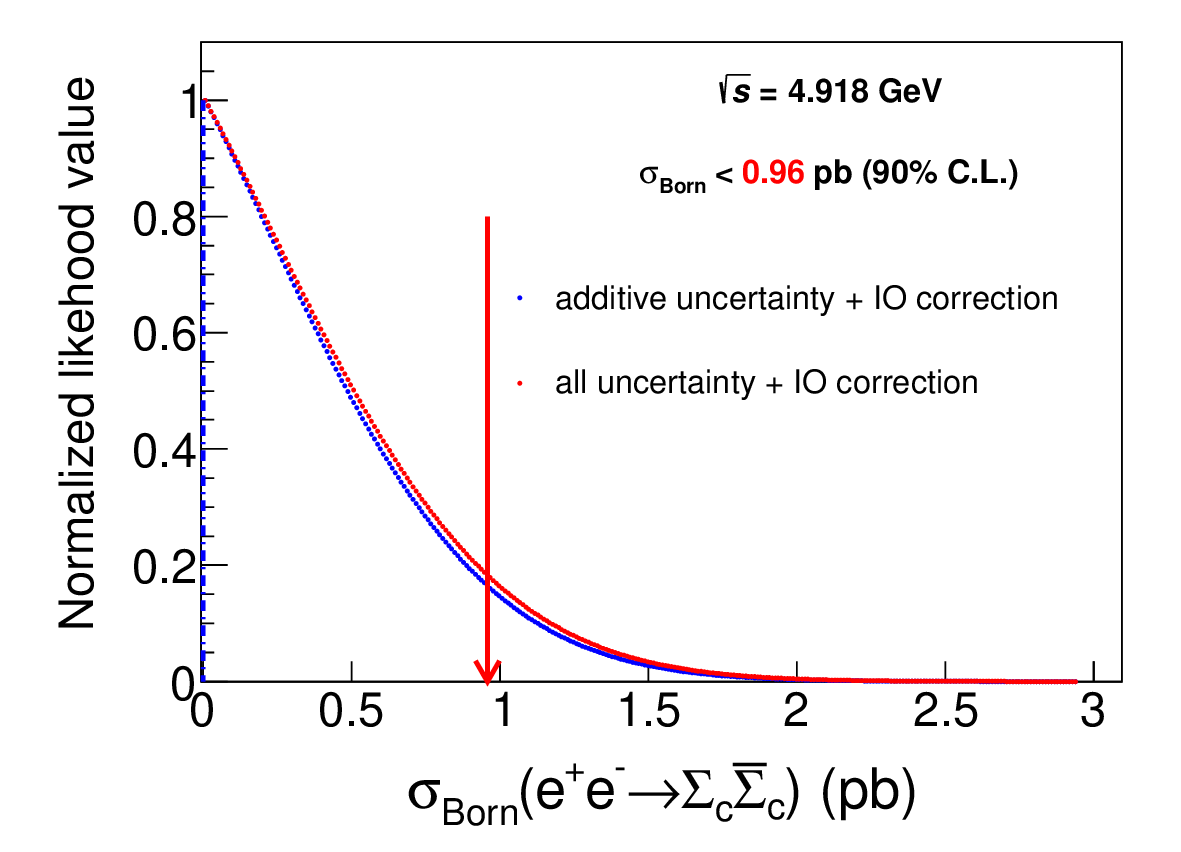}
\end{minipage}

\begin{minipage}{0.45\textwidth}
  \centering
  \includegraphics[width=6.2cm, height=4.6cm]{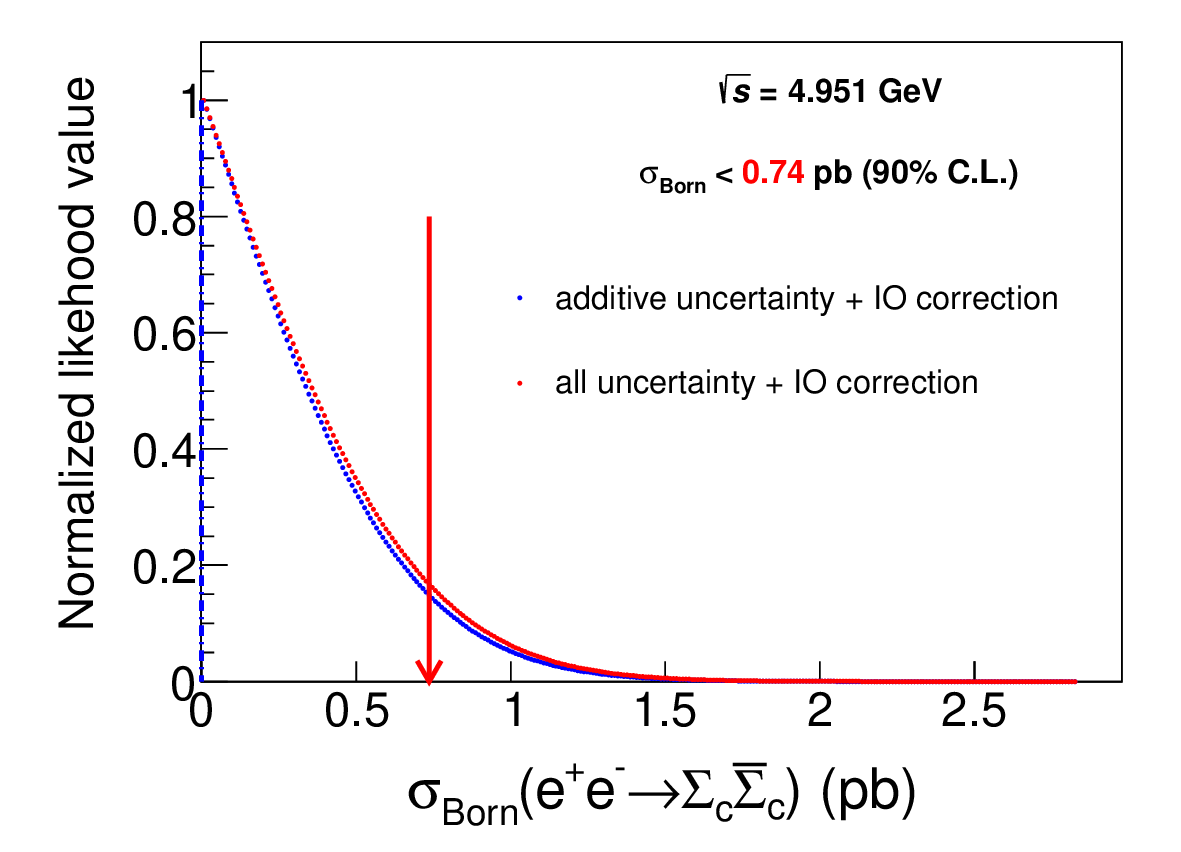}
\end{minipage}
\caption{Distributions of the normalized likelihoods in the positive region as functions of Born cross sections of $e^{+}e^{-}\to\Sigma_{c}\bar{\Sigma}_{c}$ at $\sqrt{s}$ = 4.918 GeV (top) and $\sqrt{s}$ = 4.951 GeV (bottom). The ULs at the 90$\%$ C.L. are determined from the positive region. The ``IO correction" refers to the adjustment of cross section fit bias following validation through IO checks. The blue line represents the likelihood distribution that accounts for additive systematic uncertainty and IO correction. The red line represents the likelihood distribution that incorporates both additive and multiplicative systematic uncertainties, as well as the IO correction.}
\label{fig:xs_ScSc_ul_smear}
\end{figure}

The upper limits on Born cross section considering all systematic uncertainties are obtained from the likelihood distribution $\mathcal{L}(\sigma_{\rm Born}^{{\rm smear}})$, which is defined as
\begin{equation}\label{eq:UL_smear}
\mathcal{L}(\sigma_{\rm Born}^{{\rm smear}})  = \int_{}^{} \mathcal{L}(\bar{\sigma}_{\rm Born})~{\rm exp}~(-\frac{(\sigma_{\rm Born}-\bar{\sigma}_{\rm Born})^{2}}{2\Delta(\sigma_{\rm Born})^{2}})~\mathrm{d}\sigma_{\rm Born}.
\end{equation}
Here, $\bar{\sigma}_{{\rm Born}}$ represents the Born cross section, which accounts for the additive systematic uncertainty and includes a correction to adjust the mean value based on the results of the IO check; $\Delta(\sigma_{{\rm Born}})$ denotes the total multiplicative uncertainty of Born cross section, and $\sigma_{{\rm Born}}^{{\rm smear}}$ is the Born cross section considering all systematic uncertainties. 
Figure~\ref{fig:xs_ScSc_ul_smear} illustrates all the likelihood distributions. The upper limits on the Born cross sections for each of $e^{+}e^{-}\to\Sigma_{c}^{+}\bar{\Sigma}_{c}^{-}$, $\Sigma_{c}^{0}\bar{\Sigma}_{c}^{0}$ and $\Sigma_{c}^{++}\bar{\Sigma}_{c}^{--}$ at $\sqrt{s}$ = 4.918 and 4.951 GeV are determined to be 0.96 $\rm{pb}$ and 0.74 $\rm{pb}$, respectively, at the 90$\%$ C.L..

\section{Measurements of $e^{+}e^{-}\to\Lambda_c^{+}\bar{\Sigma}_{c}^{-}$ cross section}

\begin{figure*}[htbp]
\centering
\begin{minipage}{0.35\textwidth}
  \centering
  \includegraphics[width=6.1cm, height=4.6cm]{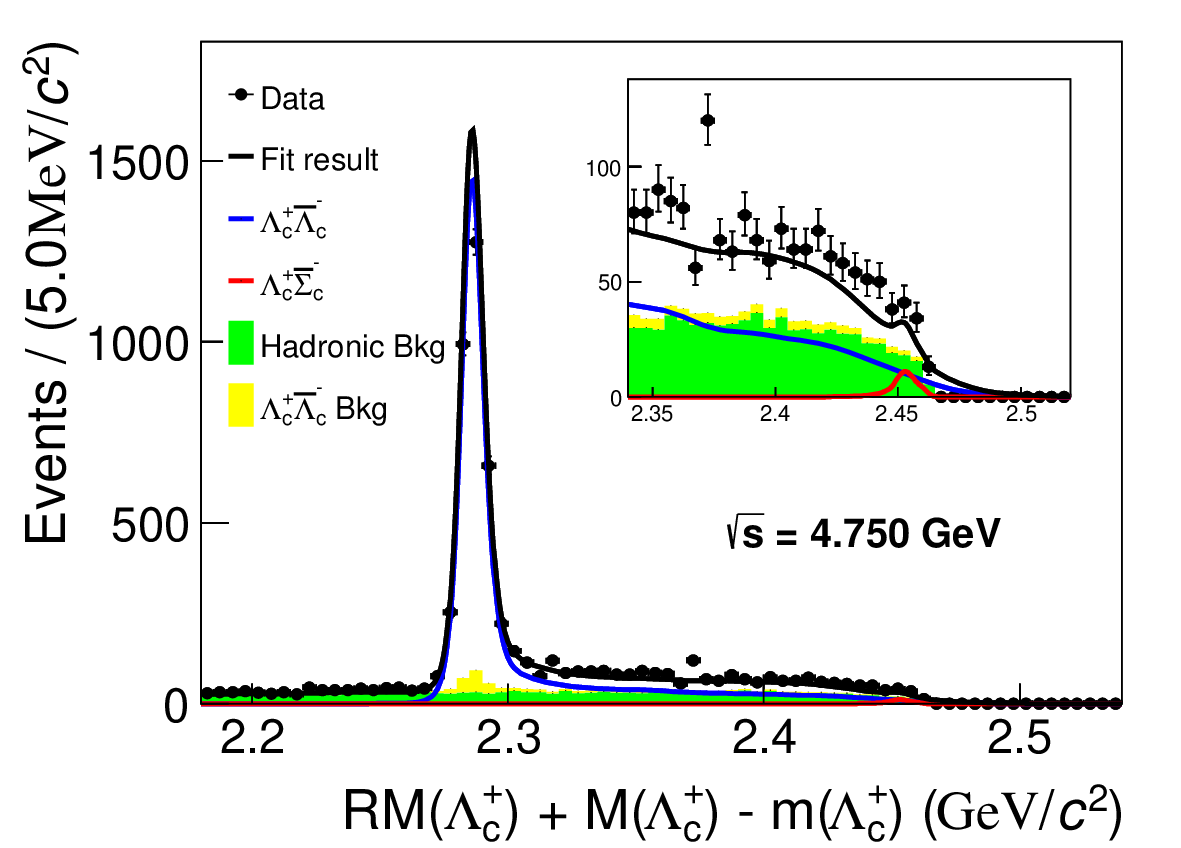}
\end{minipage}%
\hspace{0.001\textwidth}
\begin{minipage}{0.35\textwidth}
  \centering
  \includegraphics[width=6.1cm, height=4.6cm]{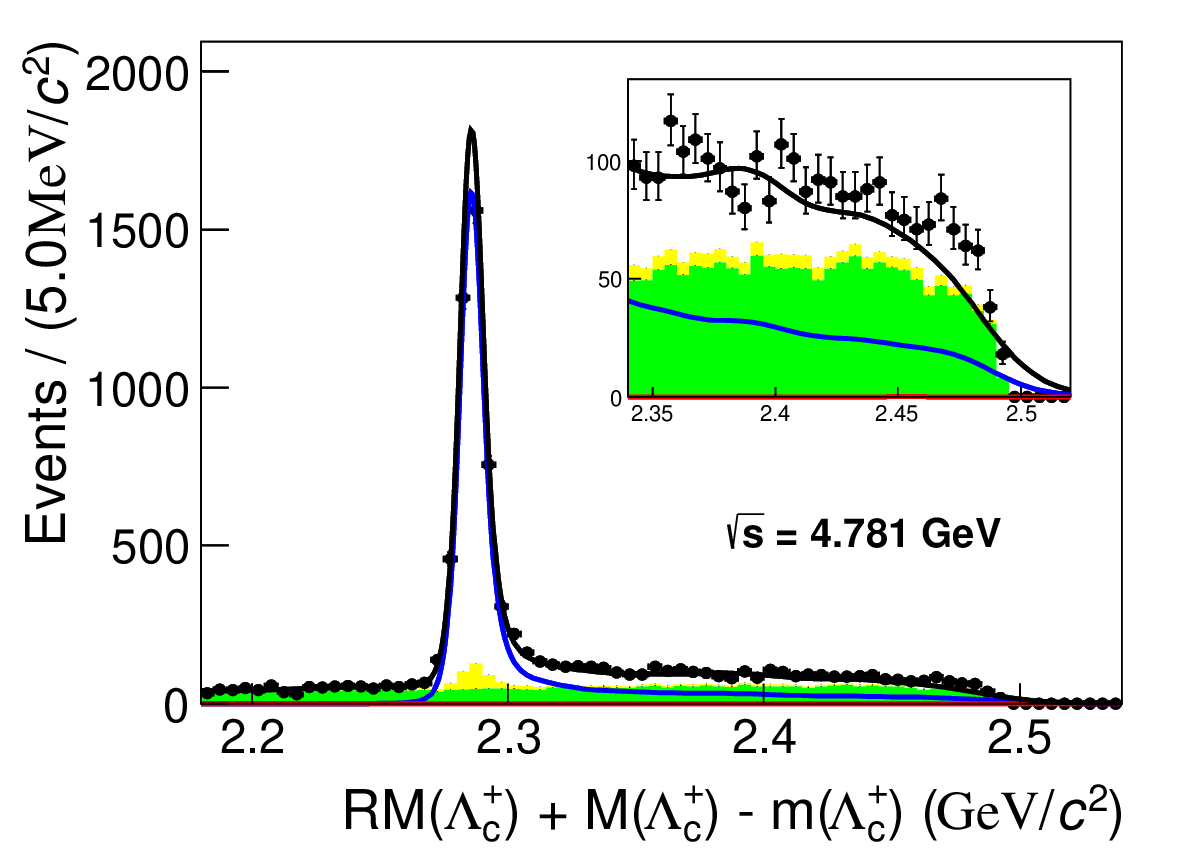}
\end{minipage}
\begin{minipage}{0.32\textwidth}
  \centering
  \includegraphics[width=6.1cm, height=4.6cm]{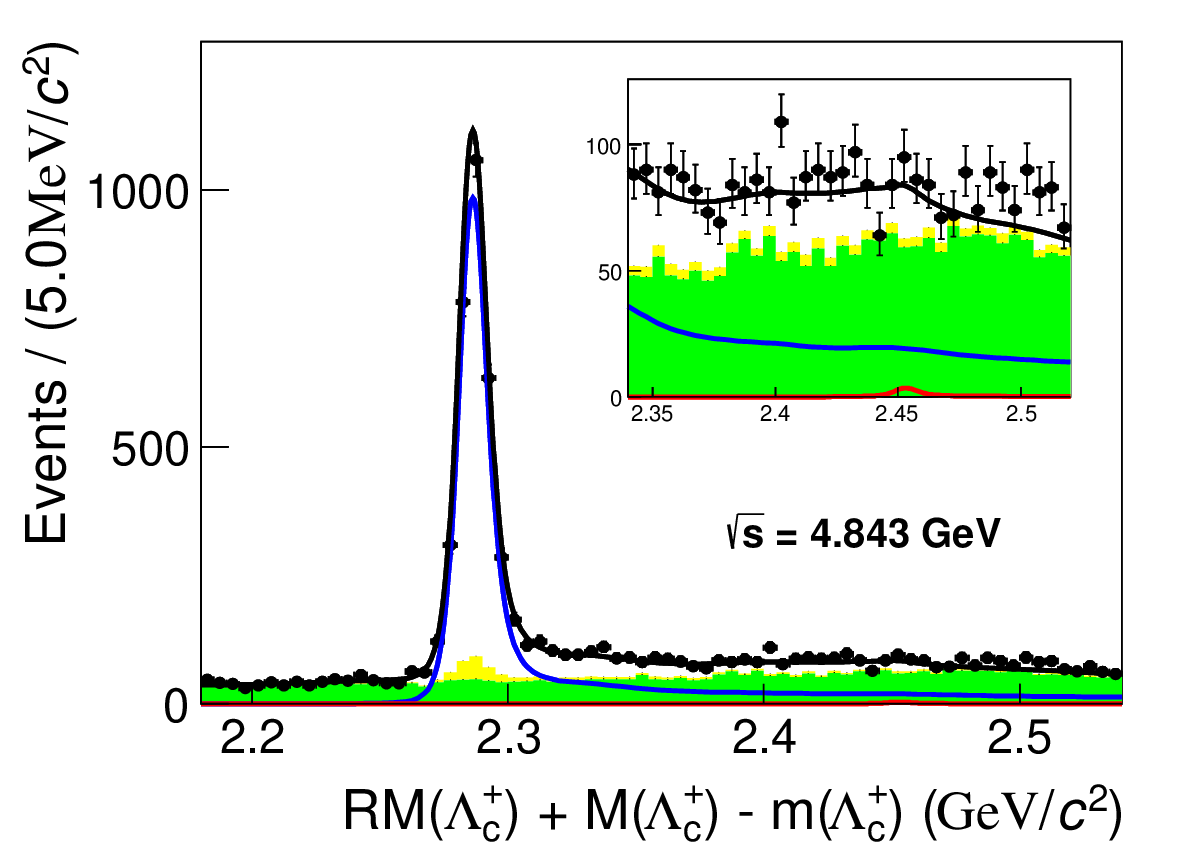}
\end{minipage}%
\hspace{0.001\textwidth}
\begin{minipage}{0.32\textwidth}
  \centering
  \includegraphics[width=6.1cm, height=4.6cm]{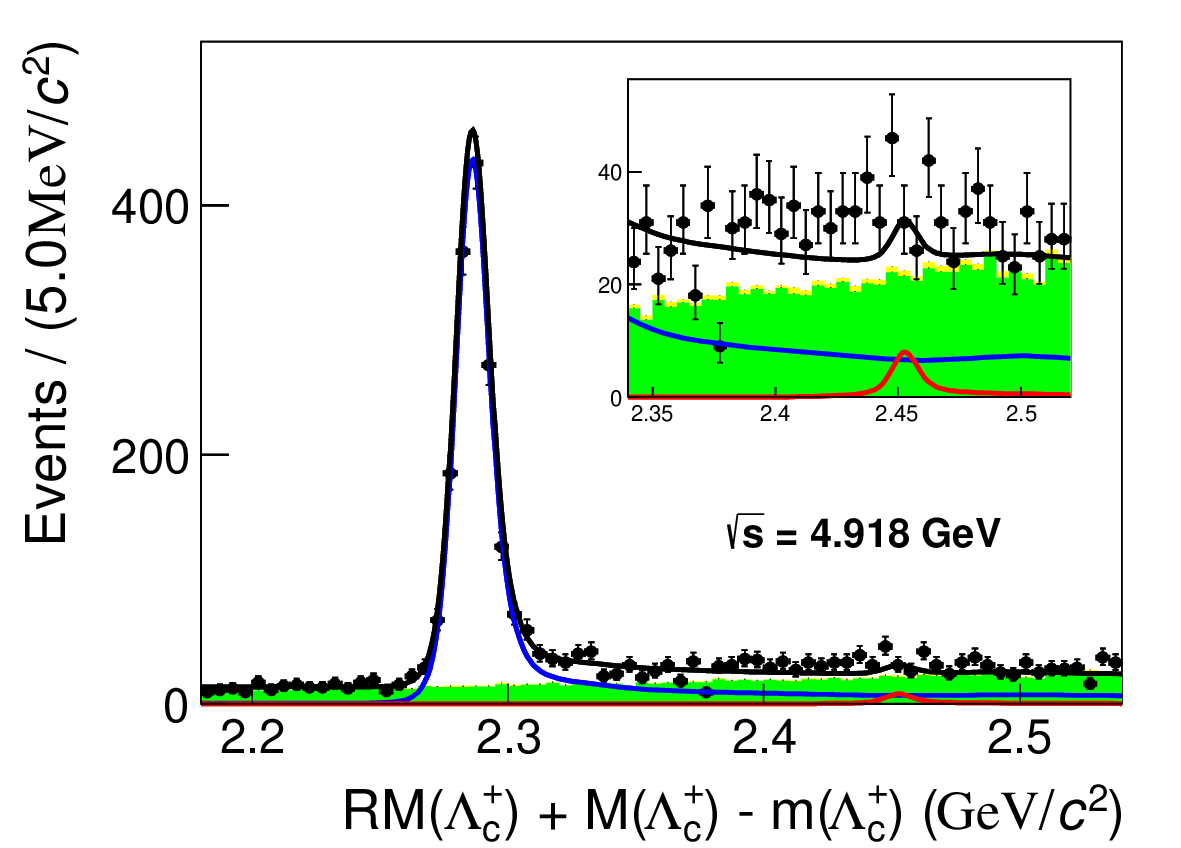}
\end{minipage}
\hspace{0.001\textwidth}
\begin{minipage}{0.32\textwidth}
  \centering
  \includegraphics[width=6.1cm, height=4.6cm]{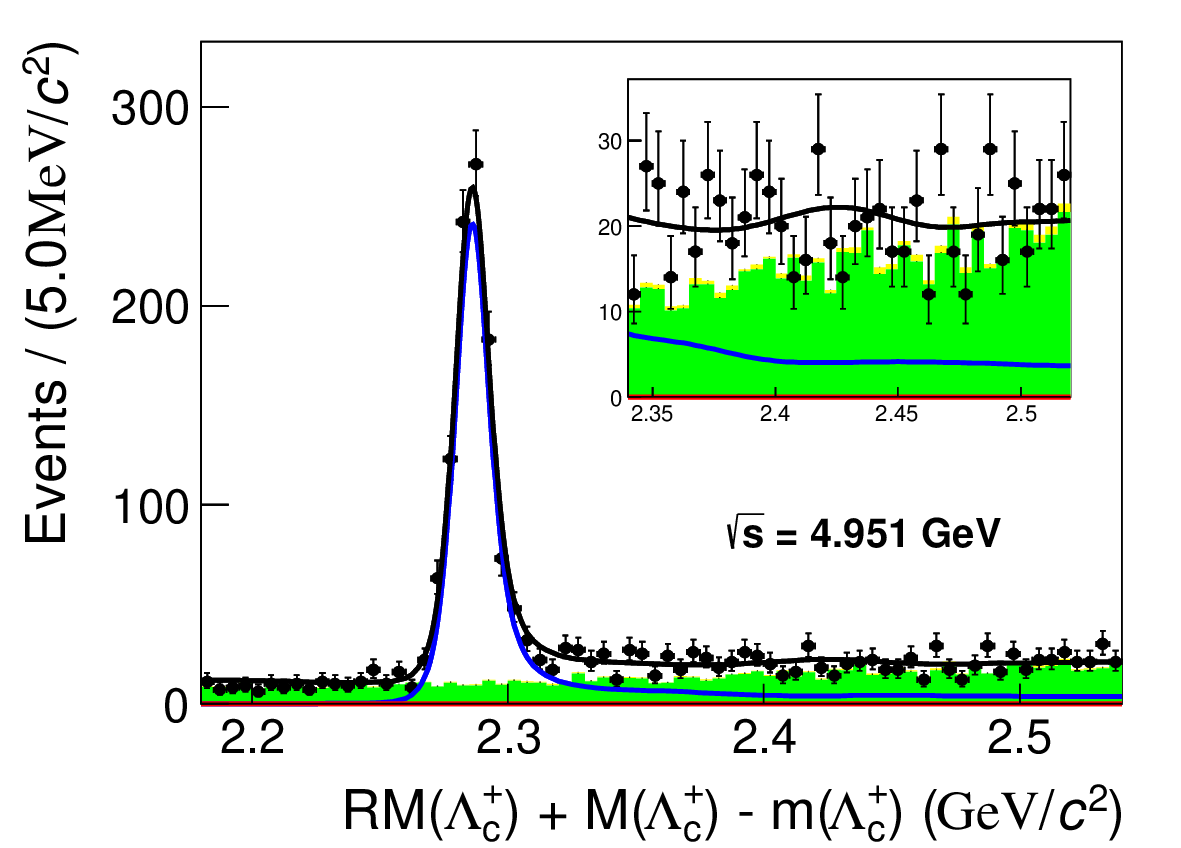}
\end{minipage}

\caption{\label{fig:xs_LcSc_nominal_fit} Simultaneous fit to the $RM(\Lambda^{+}_{c}) + M(\Lambda^{+}_{c}) - m(\Lambda^{+}_{c})$ distributions from 2.18 to 2.54$~\mathrm{GeV}/c^{2}$ in data at $\sqrt{s}=4.750$, $4.781$, $4.843$, $4.918$, and $4.951$~GeV. The small plots display the fit results in the range from 2.35 to 2.54 $\mathrm{GeV}/c^{2}$. The dots with error bars are the experimental data, while the black curves represent the total fit functions.  ``Bkg" stands for the background. }
\end{figure*}

The ``Tag $\Lambda^{+}_{c}$" method is used to measure the cross section of $e^{+}e^{-}\to \Lambda_c^{+}\bar{\Sigma}_{c}^{-}$ in the energy range from $4.750$~GeV to $4.951$~GeV. In this analysis, the spectra of $RM(\Lambda^{+}_{c}) + M(\Lambda^{+}_{c}) - m(\Lambda^{+}_{c})$ are studied in the range $(2.18,2.54)~\mathrm{GeV}/c^{2}$. 
Among all the energy points, clear $\Lambda_c^{+}\bar{\Lambda}_{c}^{-}$ signals are observed, while there is no evidence of $\Lambda_c^{+}\bar{\Sigma}_{c}^{-}$ signals, as shown in 
Fig.~\ref{fig:xs_LcSc_nominal_fit}.
Hence, to determine the ULs of the cross sections of $\Lambda_c^{+}\bar{\Sigma}_{c}^{-}$ at each energy point, an unbinned simultaneous fit to the $RM(\Lambda^{+}_{c}) + M(\Lambda^{+}_{c}) - m(\Lambda^{+}_{c})$ distributions in data is performed.
Here, the cross section ratio $R(\sigma)$, which represents the ratio between the Born cross section of $e^{+}e^{-}\to\Lambda_c^{+}\bar{\Sigma}_{c}^{-}$ and 
that of $e^{+}e^{-}\to \Lambda_c^{+}\bar{\Lambda}_{c}^{-}$, is taken as the fitting parameter.
In the fit, $e^{+}e^{-}\to\Lambda_c^{+}\bar{\Sigma}_{c}^{-}$ and $e^{+}e^{-}\to\Lambda_c^{+}\bar{\Lambda}_{c}^{-}$ signal shapes are obtained from MC simulations convolved with the same Gaussian function, whose parameters are left free. 

Based on the studies of $e^{+}e^{-}\to\Lambda_c^{+}\bar{\Lambda}_{c}^{-}$ and inclusive hadronic MC samples, no peaking background is observed within the $\Sigma^{+}_{c}$ signal region. In the fit, the $\Lambda_c^{+}\bar{\Lambda}_{c}^{-}$ non-tagged background and hadronic background are modelled using MC simulation. The yields of $\Lambda_c^{+}\bar{\Lambda}_{c}^{-}$ non-tagged models are fixed, while the contributions of hadronic backgrounds are allowed to vary. 
However, the contribution of the isospin-violating three-body process $e^{+}e^{-}\to\Lambda_c^{+}\bar{\Lambda}_{c}^{-}\pi^{0}$ remains a potential factor that could influence the signal yield, and, therefore, is considered as a source of systematic uncertainty.

The fit results corresponding to the maximum likelihood values at $\sqrt{s}=4.750$, $4.781$, $4.843$, $4.918$, and $4.951$~GeV are presented in Fig.~\ref{fig:xs_LcSc_nominal_fit}. Following the same likelihood scan method discussed at Sec.~\ref{sec:xs_ScSc_simfit}, the ULs on $R(\sigma)$ at the 90\% C.L. are obtained.

With regard to systematic uncertainties, only additive uncertainties are considered, since the multiplicative uncertainties are naturally eliminated in $R(\sigma)$. The sources of additive uncertainties include the input line shape of $e^{+}e^{-}\to\Lambda_c^{+}\bar{\Sigma}_{c}^{-}$ and the contribution of $e^{+}e^{-}\to\Lambda_c^{+}\bar{\Lambda}_{c}^{-}\pi^{0}$. 
Considering the line shape of the input cross section of $e^{+}e^{-}\to\Lambda_c^{+}\bar{\Lambda}_{c}^{-}$, the uncertainty is investigated by replacing it with a plateau characterized by $\sigma \propto \frac{1}{\sqrt{s}}$.
The possible contribution of $\Lambda_c^{+}\bar{\Lambda}_{c}^{-}\pi^{0}$ is considered by incorporating it into the fit model additionally. Its shape is extracted from the corresponding MC sample, with the yield being free.
Among the above systematic variations, the most conservative UL's are taken as the final results, which are given in Table~\ref{tab:xs_LcSc_nominal_final_result}.

\begin{figure}[htbp]
\centering
\begin{minipage}{0.35\textwidth}
  \centering
  \includegraphics[width=6.2cm, height=4.6cm]{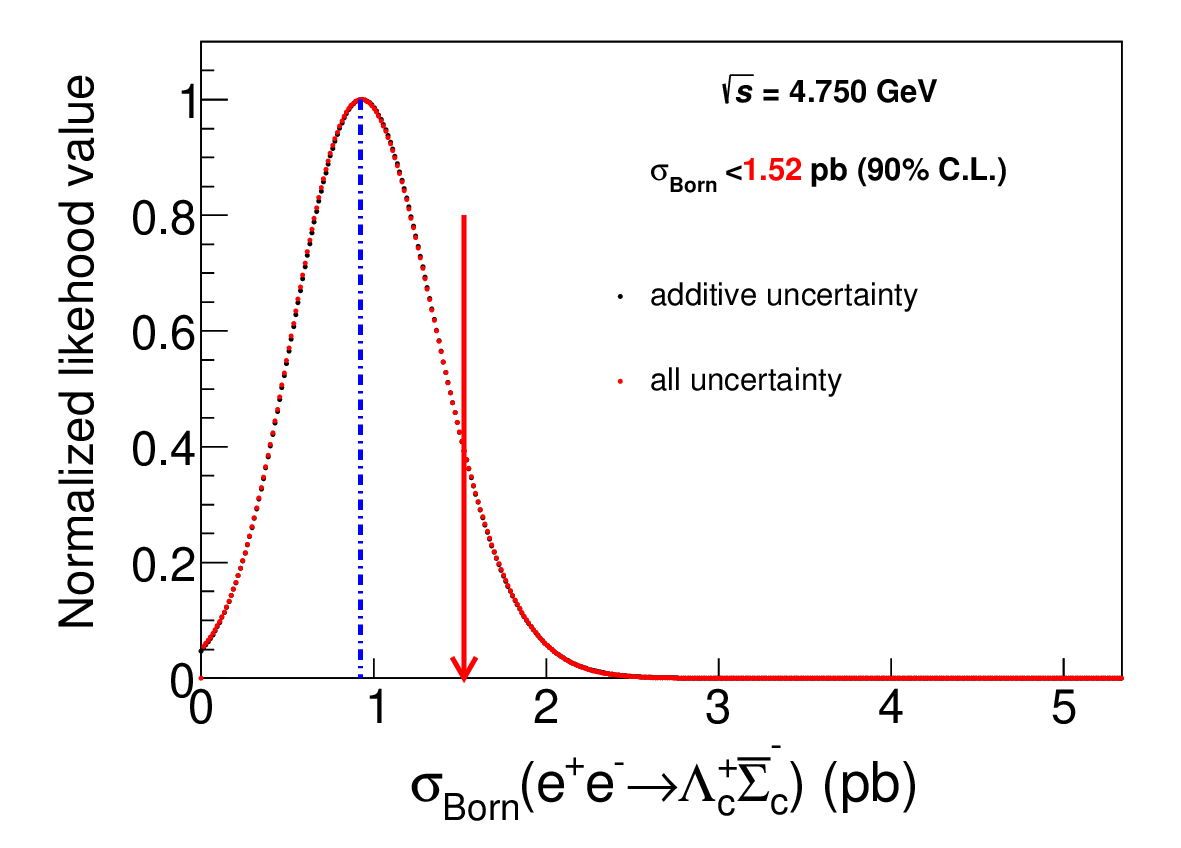}
\end{minipage}
\caption{Distribution of the normalized likelihood in the positive region as a function of Born cross section of $e^{+}e^{-}\to\Lambda_c^{+}\bar{\Sigma}_{c}^{-}$ at $\sqrt{s}$ = 4.750 GeV. The black line represents the original likelihood distribution, while the red line represents the distribution considering all uncertainties in the $e^{+}e^{-}\to\Lambda_c^{+}\bar{\Lambda}_{c}^{-}$ cross section measurement~\cite{cit:Bes_LcLc_xs_new}.}
\label{fig:xs_LcSc_ul_smear}
\end{figure}

Concerning the Born cross section $\sigma_{{\rm Born}}(e^{+}e^{-}\to\Lambda_c^{+}\bar{\Sigma}_{c}^{-})$, the likelihood function is further smeared according to Eq.~\ref{eq:UL_smear} with the input uncertainty of the Born cross section $\sigma_{\rm Born}(e^{+}e^{-}\to\Lambda_c^{+}\bar{\Lambda}_{c}^{-})$, which is studied by BESIII~\cite{cit:Bes_LcLc_xs_new}. 
The likelihood ratio distribution at $\sqrt{s}=4.750$~GeV is illustrated in Fig.~\ref{fig:xs_LcSc_ul_smear}, and the numerical results of all data samples are summarized in Table~\ref{tab:xs_LcSc_nominal_final_result}.

\begin{table*}[htbp]
\begin{center}
\caption{ Summary of the upper limits on Born cross sections of $e^{+}e^{-}\to \Lambda_c^{+}\bar{\Sigma}_{c}^{-}$ at 90\% C.L.. The results of $\sigma_{{\rm Born}}(e^{+}e^{-}\to\Lambda_c^{+}\bar{\Lambda}_{c}^{-})$ are input from Ref.~\cite{cit:Bes_LcLc_xs_new}, where the first uncertainty represents statistical uncertainty and the second one represents systematic uncertainty.\label{tab:xs_LcSc_nominal_final_result}}
\begin{ruledtabular}
\begin{tabular}{cccccc}
$\sqrt{s}$ & 4.750 GeV & 4.781 GeV & 4.843 GeV & 4.918 GeV & 4.951 GeV \\
\midrule
$R(\sigma)$ (\%) & $<1.1$ & $<0.6$ & $<1.5$ & $<3.4$ & $<1.6$\\
$\sigma_{{\rm Born}}(e^{+}e^{-}\to\Lambda_c^{+}\bar{\Lambda}_{c}^{-})$ ($\rm{pb}$) & $134\pm3\pm4$ & $127\pm2\pm4$ & $83\pm2\pm3$ & $96\pm3\pm4$ & $88\pm4\pm3$\\
$\sigma_{\rm Born}(e^{+}e^{-}\to\Lambda_c^{+}\bar{\Sigma}_{c}^{-})$ ($\rm{pb}$) & $<1.52$  & $<0.76$ & $<1.26$ & $<3.26$ & $<1.38$ \\
\end{tabular}
\end{ruledtabular}
\end{center}
\end{table*}

\section{Summary}

Using the data collected with the \mbox{BESIII} detectors, the measurement of Born cross section of $e^{+}e^{-}\to\Sigma_{c}\bar{\Sigma}_{c}$ ($\Sigma_{c}^{+}\bar{\Sigma}_{c}^{-}$, $\Sigma_{c}^{0}\bar{\Sigma}_{c}^{0}$ and $\Sigma_{c}^{++}\bar{\Sigma}_{c}^{--}$) at $\sqrt{s}$ = 4.918 and 4.951 GeV is conducted, no significant signals are observed. The upper limits at the 90$\%$ confidence level on Born cross section are estimated to be 0.96 pb and 0.74 pb under the Hypothesis 2 lineshape presented in Fig.~\ref{fig:xs_ScSc_Born_xs_Line}, respectively. 

Comparing the cross section of \( e^+e^- \to \Lambda_c^+ \bar{\Lambda}_c^- \) near the threshold (\(\sim\)250 pb)~\cite{cit:Bes_LcLc_xs_old}, a suppression of at least two orders of magnitude is observed. This suppression might be attributed to the mass difference between ``good" and ``bad" diquarks, though it remains unclear whether such a mass difference alone can account for the observed effect. 

A similar suppression is observed in measurements by the Belle collaboration at 10.58 GeV for the inclusive processes \( e^+e^- \to \Lambda_c^+ X \) and \( e^+e^- \to \Sigma_c^0 X \), where the cross sections differ by about two orders of magnitude (141.79 pb vs. 7.963 pb)~\cite{cit:Belle_inclusive_xs}. While this aligns with the trend near the threshold, it raises a new question: the cross section of \( e^+e^- \to \Sigma_c \bar{\Sigma}_c \) far from the threshold is roughly ten times larger than the upper limit value near the threshold, calling for further investigation.

In contrast, such cross section suppression is not observed in hyperon pair production. For example, the cross sections for \( \Lambda \Lambda \) pairs (\(\sim\)90 pb)~\cite{cit:lmdlmd_xs} and \( \Sigma^0 \Sigma^0 \) pairs (\(\sim\)30 pb)~\cite{cit:sigma0sigma0_xs} are comparable, indicating a distinct production mechanism compared to charmed baryons. The different behavior in hyperon production and charmed baryon productions implies the presence of certain mechanisms that have yet to be fully understood.

Regarding the isospin violating process \mbox{$e^{+}e^{-}\to\Lambda_c^{+}\bar{\Sigma}_{c}^{-}$}, the upper limit on Born cross sections at $\sqrt{s}$ = 4.750, 4.781, 4.843, 4.918 and 4.951 GeV at the 90$\%$ confidence level are reported, which are about 1\% relative to the Born cross sections of $e^{+}e^{-}\to\Lambda_c^{+}\bar{\Lambda}_{c}^{-}$. Despite the statistical constraints, this result is valuable in understanding the magnitude of electromagnetic process in the production of charmed baryons.

Due to limited statistics, no signals of $e^{+}e^{-}\to \Sigma_{c} \bar{\Sigma}_{c}$ or $e^{+}e^{-}\to\Lambda_c^{+}\bar{\Sigma}_{c}^{-}$ have been observed. In coming years, improved measurements are expected in the upgrade of the current BEPCII~\cite{cit:BEPCII-U}, which would have threefold instant luminosity, and the future super $\tau$-charm facility~\cite{cit:STCF} will greatly advance the studies.

\begin{acknowledgments}
The BESIII Collaboration thanks the staff of BEPCII (https://cstr.cn/31109.02.BEPC) and the IHEP computing center for their strong support. This work is supported in part by National Key R\&D Program of China under Contracts Nos. 2023YFA1606000, 2023YFA1606704; National Natural Science Foundation of China (NSFC) under Contracts Nos. 11635010, 11735014, 11935015, 11935016, 11935018, 12025502, 12035009, 12035013, 12061131003, 12192260, 12192261, 12192262, 12192263, 12192264, 12192265, 12221005, 12225509, 12235017, 12361141819; the Chinese Academy of Sciences (CAS) Large-Scale Scientific Facility Program; the CAS Center for Excellence in Particle Physics (CCEPP); Joint Large-Scale Scientific Facility Funds of the NSFC and CAS under Contract No. U1832207; CAS under Contract No. YSBR-101; 100 Talents Program of CAS; The Institute of Nuclear and Particle Physics (INPAC) and Shanghai Key Laboratory for Particle Physics and Cosmology; German Research Foundation DFG under Contract No. FOR5327; Istituto Nazionale di Fisica Nucleare, Italy; Knut and Alice Wallenberg Foundation under Contracts Nos. 2021.0174, 2021.0299; Ministry of Development of Turkey under Contract No. DPT2006K-120470; National Research Foundation of Korea under Contract No. NRF-2022R1A2C1092335; National Science and Technology fund of Mongolia; National Science Research and Innovation Fund (NSRF) via the Program Management Unit for Human Resources \& Institutional Development, Research and Innovation of Thailand under Contract No. B50G670107; Polish National Science Centre under Contract No. 2019/35/O/ST2/02907; Swedish Research Council under Contract No. 2019.04595; The Swedish Foundation for International Cooperation in Research and Higher Education under Contract No. CH2018-7756; U. S. Department of Energy under Contract No. DE-FG02-05ER41374
\end{acknowledgments}

\appendix

\nocite{*}

\bibliography{apssamp}

\end{document}